\documentclass[draft]{agujournal2019}
\usepackage{url} 
\usepackage[inline]{trackchanges} 
\usepackage{soul}
\usepackage{amsmath}
\usepackage{lscape}
\usepackage{enumitem}
\draftfalse

\journalname{JGR: Planets}

\begin{document}

%
%


\title{Birth and decline of magma oceans.\\ Part 2: wobbling thermal history of early accreted planetesimals.}

%
%




\authors{Cyril Sturtz\affil{1}, Angela Limare\affil{1}, Stephen Tait\affil{1} and \'Edouard Kaminski\affil{1}.}

 \affiliation{1}{Universit\'e de Paris, Institut de Physique du Globe de Paris, CNRS, F-75005, France}





\correspondingauthor{Cyril Sturtz}{sturtz@ipgp.fr}




\begin{keypoints}
	\item We model the formation of magma oceans in small planetesimals and their evolution during crystallization.
	\item Incomplete segregation of suspended crystals from convecting magma oceans results in layered planetesimals mantles.
	\item The presence of a flotation crust induces cycles of remelting and crust destabilization that produce in turn a non monotonic thermal evolution.	
\end{keypoints}

%
%

%
%

\begin{abstract}
\indent A theoretical model that describes the evolution of a suspension in which crystals can sediment to form a dense cumulate or may produce a light flotation crust has been derived in a companion paper. We use this model to study the thermal history of early accreted planetary bodies accreted during the very early stages of the formation of the solar system. We study the conditions required to form and preserve flotation crusts and basal cumulates, and the implications for the thermal evolution of planetesimals. We calculate the temperature evolution in an early accreted planetesimals internally heated by the decay of $\rm{^{26}Al}$ and $\rm{^{60}Fe}$. For planetesimal with radius $R>30\, \rm{km}$, partial melting reaches 40\%, planetesimals undergo a rheological transition and form a magma ocean, i.e.: a suspension from which crystals can segregate and form a floating crust and/or a dense cumulate. Because of the formation of an insulation floating crust, this magma ocean episode is characterized by a relatively long time life, a slow cooling rate, and a weak surface heat flux. The model further predicts a cyclic evolution where episodes of crustal thickening alternate with episodes of melting-induced crustal thinning. These cycles produce in turn an oscillating thermal history and prevent runaway thermal heating of the planetesimals. At the end of the magma ocean episode, when the fraction of crystals becomes larger than 60\%,  a transition occurs to solid-state convection in the planetesimal's mantle. This stage is characterized by high viscous shear stress that tends to erode previously formed crystal deposits. However, the time scale of this erosion process is larger that the lifetime of the planetesimal. Hence solid-state evolution can be described by a well mixed mantle embedded by a metastable crust and a preserved cumulate at its base.
\end{abstract}

\section*{Plain Language Summary}
\indent Planetary formation occurred during the first million years after the Sun formation. In this period, cold dust and gas accreted to form planetesimals, which themselves accreted later to form planets. When planetesimals are large enough, they undergo partial melting due to heat produced by radioactive decay of extinct elements. The degree of melting can be high enough for core differentiation to occur. Differentiation produces a molten proto-mantle carrying crystals, which behaves like a convective suspension and is called a magma ocean. In this study, we illustrate how crystals can segregate from the liquid suspension to produce a flotation crust or a basal cumulate during this period. We find that crystals heavier than the liquid sediment to form dense cumulates whether crystals lighter than the fluid form a flotation crust. Crystal segregation is however incomplete and the crust and the cumulate embed a well mixed mantle. The flotation crust has an insulation effects on the planetesimal and decreases the rate at which the internal heat produced by radiogenics is evacuated. As a consequence, the system undergoes a long and moderately hot episode of magma ocean where the crust adopts a cyclic evolution of flotation-induced crustal thickening alternates with melting-induced crustal thinning.

%
%

\section{Introduction}

	\indent Planetary accretion occured in the first million years (Myr) after the formation of calcium-aluminium inclusion (CAI) \cite{Baker05,Morbidelli16}. During this period, undifferentiated cold pebbles accreted to form planetesimals that further partially melted and differentiated \cite{Kruijer14}. The degree of melting in the rocky undifferentiated body is the key parameter in the planetesimal evolution, and is intimately linked to the heating by radioactive decay. In particular $^{26}\rm{Al}$ and, to a much smaller extent $^{60}\rm{Fe}$, produced a temperature increase that can be large enough to induce partial melting \cite{Neumann12}, and the formation of a metallic core by planetary-scale differentiation \cite{Hoink06,Sahijpal07}. $\rm{^{26}Al}$ then partitioned into the silicate mantle where temperature continues to rise and can reach the solidus. In large enough planetesimals, the temperature increase can be close to the liquidus and episodes of magma oceans may occur.\\
	\indent  A magma ocean involves a mixture of silicate liquid and crystals that rheologically behaves like a fluid \cite{Taylor92,ET12}. As long as the fraction of crystal is smaller than 60\%, the magma ocean convects like a viscous suspension \cite{Solomatov00}. In this regime, crystals can leave the suspension \cite{Lavorel09} to form a flotation crust or a basal cumulate according to their buoyancy. The segregation of crystals induces eventually a petrological evolution of the magma ocean, and can produce to a differentiated mantle.\\
	\indent The chemical evolution of a magma ocean is usually described using petrological models coupled with simple segregation criteria based only on the buoyancy of crystals. Heavy crystals form cumulates of dunites and/or harzburgites at the core-mantle boundary (CMB) by sedimentation \cite{Righter97}, whereas light crystals accumulate at the surface, like plagioclases that formed the anorthosite crust of the Moon \cite{Wood70a,Wood70b,Warren85}. More complex scenarios can be proposed to account for the petrological diversity of rocks encountered in asteroids. For instance, \citeA{Mandler13} used a two-steps model to explain the formation of heavy diogenites at the surface of asteroid 4 Vesta through the crystallization of magma chambers that formed after the solidification of an eucritic crust. However none of these scenarios encompasses the bulk dynamics and complete thermo-chemical evolution of the magma ocean and the associated thermal histories are bound to remain monotonic. Moreover, although such models can provide important clues on the evolution of magma oceans, they can not account for the complex evolution evidenced in some asteroids. In the case of Vesta for instance, \citeA{Barrat14} concluded that the rare earth elements (REE) abundances in diogenites display a discrepancy that can not be explained by the monotonic crystallization of a single parent magma for the howardites-eucrites-diogenites (HED) series. The interpretation of REE composition of pallasites, iron meteorites containing olivine crystals, also suggests remelting episodes, hence a non monotonic thermal evolution \cite{Barrat21} and remelting episodes of cumulates that have previously crystallized \cite{Yamaguchi09,Yamaguchi11}.\\
	\indent The aim of this paper is to show how a global modeling of thermal evolution of a magma ocean, including the modeling of the behavior of crystals in the magmatic suspension, can provide a direct explanation of complex petrological features in asteroids. To that aim, we use the erosion/deposition laws established theoretically and tested experimentally in a companion paper \cite{Sturtz21b}. First, we describe the scenarios that lead to the formation of a magma ocean in a planetesimal. Then, we study the formation of cumulates or flotation crusts by crystal/melt segregation from a convecting magma in planetesimals. We finally model the complete thermal history based on a crystallization toy model for early accreted planetesimals that provides a framework to assess and predict (i) the condition to form crystal layers, (ii) their formation time-scales, (iii) whether they can be preserved or not. 
	
%
%
	
\section{From early accretion to magma ocean episode}
	
	\indent Planetary formation models suggest that planetesimals formed through the accretion of cold chondritic pebbles during the first few Myr after CAI. Thermal evolution is linked to the accretion history through two important key parameters: the time at which accretion begins, and the planetesimal size. In the following, we consider planetesimals that accrete early ($t=0$ Myr) and we study their thermal evolution as a function of their radius only.

%
%

	\subsection{Petrological composition}
\begin{figure}
	\centering
	\includegraphics[width=0.55\textwidth]{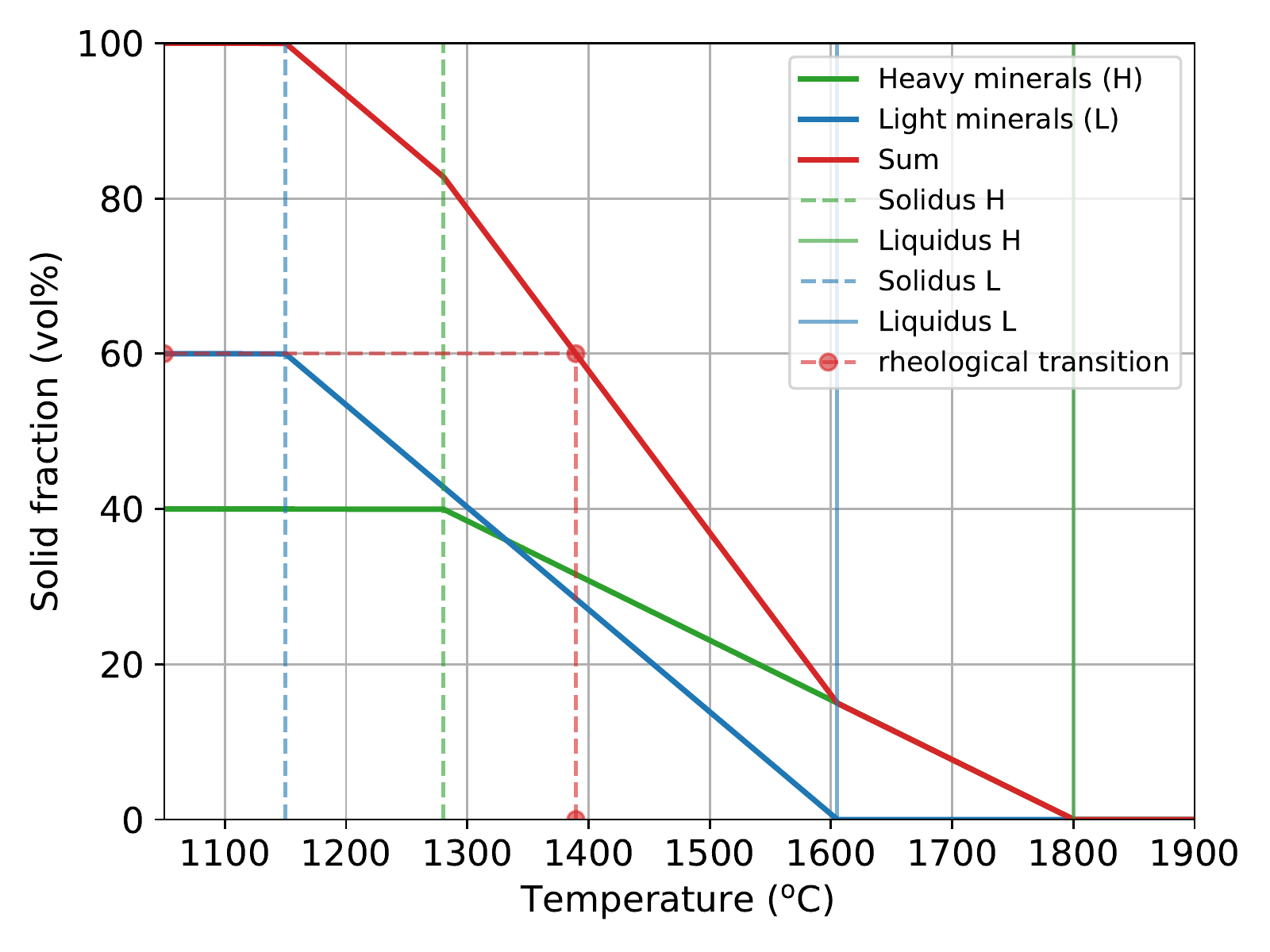}
	\caption{Solid fraction in the silicate part of a planetesimal as a function of temperature. The silicate liquid can produce two minerals: a positively buoyant one (the light one, in blue), and a negatively buoyant one (the heavy one, in green). The red curve gives the total solid content of the system. We also display the rheological transition, defined by a total solid content of $60\rm{vol}\%$, which corresponds the temperature $T_{\rm{RT}}=1390\rm{^oC}$.}
	\label{fig:CrystSerie}
\end{figure}

	\indent We consider a spherical undifferentiated planetesimal of radius $R$ with a relative mass fractions $x_{\rm{Fe}}$ of iron and $x_{\rm{Sil}}$ of silicate. The mean density of the planetesimal $\rho$ is given by:
\begin{linenomath*}
	\begin{equation}
		\frac{1}{\rho}=\frac{x_{\rm{Fe}}}{\rho_{\rm{Fe}}}+\frac{x_{\rm{Sil}}}{\rho_{\rm{Sil}}},
	\end{equation}
\end{linenomath*}	
	with $\rho_{i}$ the density of $i=\rm{Fe}$ and $\rm{Sil}$.\\
	\indent For the the composition of the silicate material, we consider a toy-model in order to study the formation of either a flotation crust and a cumulate. In the model, the silicate material is composed of two main minerals, a heavy one (indexed $H$ hereafter) negatively buoyant, and a lighter one (indexed $L$ hereafter), supposed to be positively buoyant when it nucleates in the parent magma. We assume that both crystals are present in the silicate part in proportion $\phi_{\rm{L},0}/\phi_{\rm{H},0}=60/40$, following the ratio between pyroxene and olivine measured in some chondrites (e.g.: \citeA{Righter97}). The evolution of the solid content in the silicate part is displayed in Figure \ref{fig:CrystSerie} and the total solid content is given by the relation:
\begin{linenomath*}
	\begin{eqnarray}
		\phi&=&\phi_{\rm{L},cr}+\phi_{\rm{H},cr},\label{eq:phi}\\
		\phi_{\rm{L},cr}&=&\phi_{\rm{L},0}\, \frac{T_l^{\rm{L}}-T}{T_l^{\rm{L}}-T_s^{\rm{L}}}, \label{eq:phiLcr}\\
		\phi_{\rm{H},cr}&=&\phi_{\rm{H},0}\, \frac{T_l^{\rm{H}}-T}{T_l^{H}-T_s^{H}}, \label{eq:phiHcr}
	\end{eqnarray}
\end{linenomath*}	
where $T$ is the bulk temperature, $T_l^i$ and $T_s^i$ are the liquidus and the solidus temperature of mineral $i$ respectively, and $\phi_{i,0}$ its initial proportion (in vol$\%$).

%
%
	\subsection{Conduction regime}
	\indent Once accreted, the planetesimal is initially at the protoplanetary disk temperature ($T\approx 0\rm{^oC}$) and undifferentiated. Its thermal evolution is given by the energy budget:
\begin{linenomath*}
	\begin{equation}
		\rho c_p \frac{\rm{d} T}{\rm{d} t }= H - \frac{3}{R}\, Q_s, \label{eq:cond}
	\end{equation}
\end{linenomath*}		
	where $c_p$ is the average heat capacity of the planetesimal, $T$ its volume average temperature, $H$ the rate of internal heating, $R$ the planetesimal radius, and $Q_s$ the surface heat flux. \\
	\indent We consider that $^{26}\rm{Al}$ and $^{60}\rm{Fe}$ are the two radiogenic elements active during this episode and the rate of internal heating in the undifferentiated planetesimal  is given by chondritic abundances:
\begin{linenomath*}
	\begin{eqnarray}
		H_{0,\rm{Fe}}&=& \rho \, F_{\rm{Fe}} \left[\frac{^{60}\rm{Fe}}{^{56}\rm{Fe}}\right]_0\, \frac{\mathcal{E}_{\rm{Fe}}}{t_{1/2}^{\rm{Fe}}}\, e^{-t/t_{1/2}^{\rm{Fe}}}, \label{eq:Fe}\\
		H_{0,\rm{Al}}&=& \rho \, F_{\rm{Al}} \left[\frac{^{26}\rm{Al}}{^{27}\rm{Al}}\right]_0\, \frac{\mathcal{E}_{\rm{Al}}}{t_{1/2}^{Al}}\, e^{-t/t_{1/2}^{\rm{Al}}}, \label{eq:Al}
	\end{eqnarray}
\end{linenomath*}	
where $F_{j}$ is the chondritic abundance of $j$, $[j/j]_0$ is the initial abundance of nuclides $j$, $\mathcal{E}_{j}$ is the decay energy per atom, $t_{1/2}^{j}$ the half life. The values used are displayed in Table \ref{tab:paramsMGP}. The surface heat flux $Q_s$ is given by the analytical solution of the heat equation for a sphere that cools by conduction (\citeA{CarslawJeager}, pp. 245). \\
\indent The resulting thermal history shows that the bulk average temperature increases from zero to a maximum (peak) temperature before decreasing again to zero, as illustrated in \ref{App:SLC}. These peak temperatures reached in the conductive regime are shown in Figure \ref{fig:CondConv} as a function of the planetesimal size. One can note that conduction implies unrealistic temperatures that predict vaporization of planetary bodies larger than 100 km. This unrealistic conclusion can be corrected if more efficient convective heat transfer is taken into account.

%
%
	\subsection{Onset of solid-state convection}
\begin{figure}
	\centering
	\includegraphics[width=0.6\textwidth]{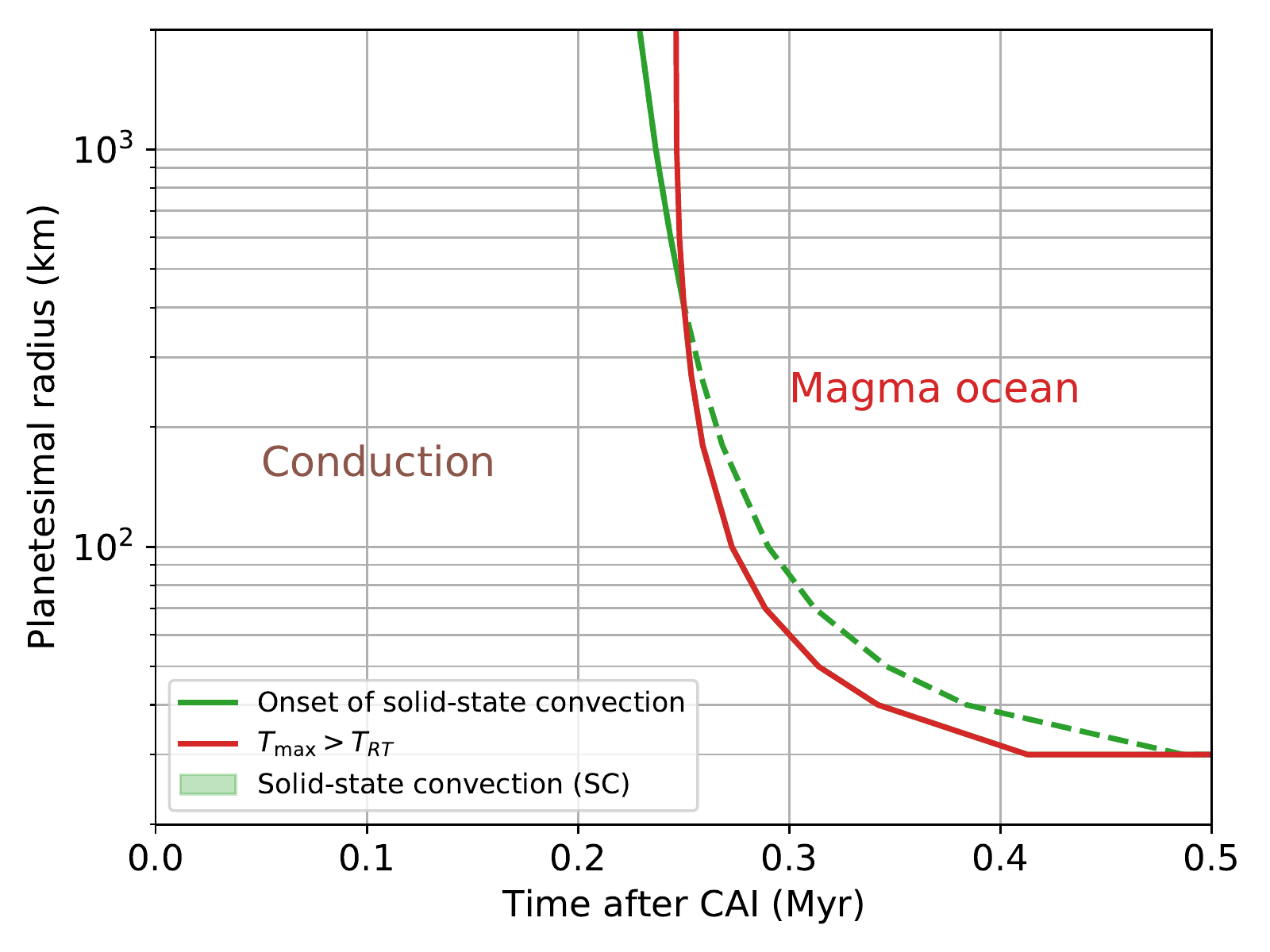}
	\caption{Regime diagram showing the thermal regime as a function of the planetesimal radius. Three types of evolution appear: (i) purely conductive regime and undifferentiated body ($R<30\, \rm{km}$), (ii) planetesimals that abruptly transition from the conductive regime directly to magma ocean ($30\ \rm{km}<R<400\ \rm{km}$), (iii) planetesimals that experience a short solid-state convection episode between the conductive regime and the magma ocean episode ($R>400\ \rm{km}$).}
	\label{fig:SSC_MO}
\end{figure}

\begin{figure}
	\centering
	\includegraphics[width=0.6\textwidth]{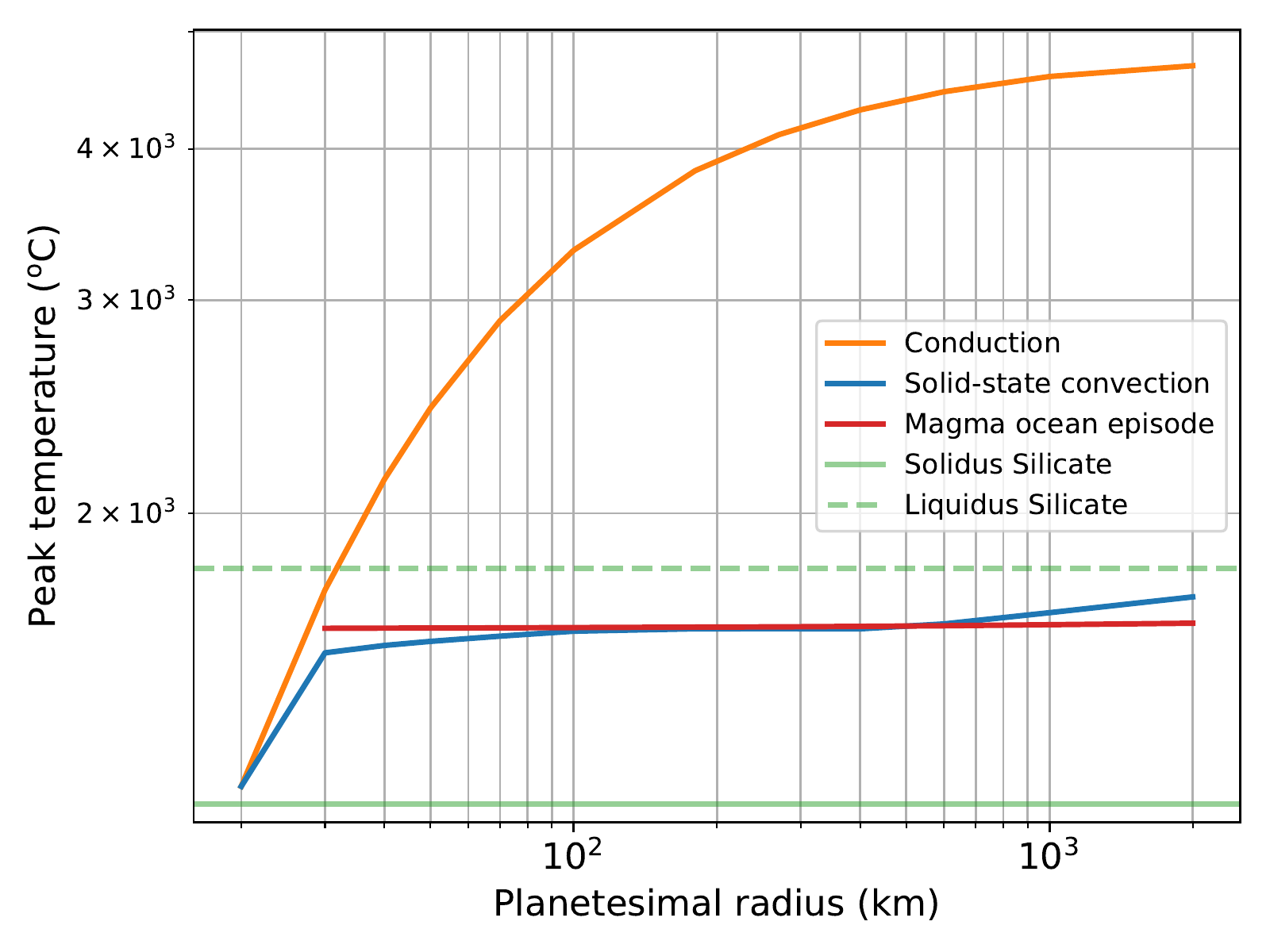}
	\caption{Peak temperature reached  in early accreted planetesimals ($t_0=0\, \rm{Myr}$), heated by radiogenic elements, as a function of their radius. Three thermal models are shown: planetesimals that transport heat only by conduction (orange), those which experience solid-state convection (blue) and those undergoing a magma ocean episode (red). Magma ocean only exists for planetesimals with $R>30\, \rm{km}$ as discussed in the text.}
	\label{fig:CondConv}
\end{figure}

	\indent As  temperature rises, the planetesimal begins to melt. \citeA{Kaminski20} showed that this melting implies a sharp decrease of viscosity and can trigger Rayleigh-Taylor instabilities, hence will trigger solid-state convection. Two characteristics number describe this internally heated convective regime \cite{Roberts67}: the Rayleigh-Roberts $Ra_H$ and the Prandtl numbers $Pr$ defined as follows:
\begin{linenomath*}
	\begin{eqnarray}
		Ra_H &=& \frac{\alpha \rho g H R^5}{\kappa \eta \lambda}, \label{eq:RaHdef}\\
		Pr      &=& \frac{\nu}{\kappa}, \label{eq:Prdef}
	\end{eqnarray}
\end{linenomath*} 	
	with $\alpha$ the average thermal expansion coefficient, $g$ the surface gravity, $H$ the rate of internal heating, $R$ the planetesimal radius, $\kappa$ the average thermal diffusivity, $\nu=\eta/\rho$ is the kinematic viscosity, $\eta$ the dynamic viscosity, and $\lambda$ the average thermal conductivity. $Ra_H$ characterizes the vigor of convection and $Pr$ quantifies the importance of inertia. The onset of convection requires two criteria. First, $Ra_H$ must be larger than the critical value $Ra_{H,c}$ which, for a spherical planetesimal with rigid boundary condition, is $Ra_{H,c}=5758$ \cite{Schubert01}. Second, in the case of strong temperature dependance of viscosity, once the critical Rayleigh-Roberts number is reached, the onset of convection is delayed and the it occurs under a stagnant lid where heat transfer occurs by conduction \cite{Davaille93,Choblet00,Kaminski20}. In the cases considered here, transition between the conductive regime and the solid-state regime occurs between 0.2 and 0.5 Myr (Figure \ref{fig:SSC_MO}). \\
	\indent Viscosity is the key parameter for convection in the partially molten mantle. It depends on two main parameters: (i) the temperature, modeled by an Arrhenius law and (ii) the melt content. If the degree of melting is low, typically smaller than $40\%$, the influence of the melt content on the viscosity can be modeled by the following law (\citeA{Kaminski20} and references herein):
\begin{linenomath*}
	\begin{eqnarray}
		\eta(T,\phi)&=&\eta_0 \, f(\phi)\, \exp \left[\frac{E_{a,\rm{Sil}}}{R_g}\left(\frac{1}{T}-\frac{1}{T_0}\right)\right], \label{eq:visco_SC1}\\
		f(\phi)&=&\exp \left[-\sigma_{\rm{Sil}} (1-\phi) \right], \label{eq:visco_SC2}
	\end{eqnarray}
\end{linenomath*} 
 where $\phi$ is the volume fraction of solid given by (\ref{eq:phi}), $R_g$ is the ideal gas constant, $\sigma_{\rm{Sil}}$ the melt constant of silicate, and we assume that the mantle viscosity at $T_0=1000\, \rm{^oC}$ is $10^{18}\ \rm{Pa\, s}$. \\
 \indent Once solid-state convection has started we take it into account in the energy budget (\ref{eq:cond}) using reference scaling laws for $Q_s$ (see \ref{App:SLC} for more details). As already discussed in \citeA{Kaminski20}, we obtain that solid-state convection avoids the catastrophic thermal state predicted by the conductive regime, and to maintain the temperature at ``moderate", sub-liquidus, values thanks to more efficient heat losses. The peak temperature reached in the planetesimal in the case of solid-state convection increases weakly as a function of the planetesimal radius (Figure \ref{fig:CondConv}). This is because the ratio between the energy lost by the surface heat flux over the heat generated by radioactive decay, proportional to the surface over volume ratio, decreases with the planetesimal radius. However, this plateau still spans from 1550 to 1700$\rm{^oC}$, which implies a melt content of 88-92 vol\%. At this degree of melt, the mush becomes a slurry and the regime of convection changes from solid-state convection to liquid-like convection.

%
%
	
	\subsection{Rheological transition: onset of magma ocean episode}
	\label{subsec:RT}
\begin{figure}
 	\centering
	\includegraphics[width=0.6\textwidth]{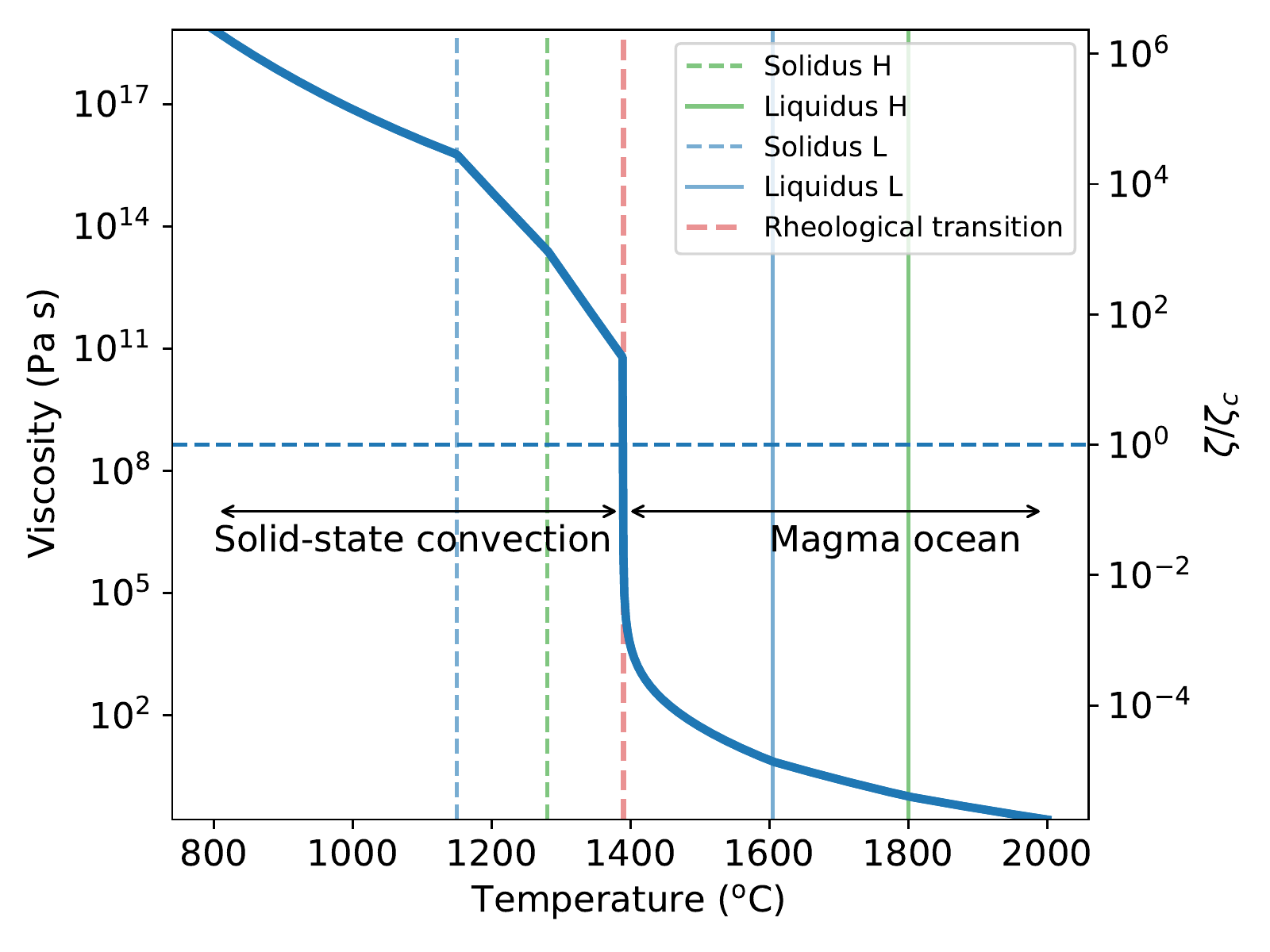}
	\caption{Viscosity function and corresponding Shields number taking into account the rheological transition, based on the crystallization series described in Figure \ref{fig:CrystSerie}. The rheological transition occurs for a solid content of $\phi=60\rm{vol}\%$, which corresponds, in our simplified two-mineral model, to the temperature $T_{\rm{RT}}=1390\rm{^oC}$. The viscosity function constrains directly the Shields number, which is calculated here for $Ra_H=10^{20}$ and parameters summarized in Table \ref{tab:paramsMGP}. The horizontal dashed blue line stands for the critical value of the Shields number $\zeta=\zeta_c$.}
	\label{fig:RT}
 \end{figure}

\indent Although moderately, temperature continues to increase in the solid-state convection regime and so does the melt fraction. Beyond a certain degree of melt, the solid matrix is not stable anymore and the mixture of melt and crystals behaves like a liquid or, more precisely, a liquid suspension. The regime of convection changes beyond the rheological transition that occurs for $\phi_{\rm{RT}}=60\%$ for spherical crystals \cite{Guazzelli18,Solomatov00}. Assuming batch melting, this critical solid content corresponds to a temperature $T_{\rm{RT}}$ that only depends on the composition of the silicate mantle of the planetesimal. In the case of our toy model, $T_{\rm{RT}}=1390\rm{^oC}$.\\
\indent  To take into account the rheological transition, we adapt the model presented by \citeA{Kaminski20} using a Krieger-Dougherty law \cite{Guazzelli18} so that $f(\phi)$ becomes:
\begin{linenomath*}
	\begin{eqnarray}
		f(\phi)&=&\left(1-\frac{\phi}{\phi_{c}}\right)^{-2.5\phi_{RT}}.\label{eq:ViscosityKD}
	\end{eqnarray}
\end{linenomath*}
To ensure the continuity of the viscosity at $\phi=\phi_{RT}$, $\phi_c$ is adjusted to get a viscosity at liquidus $\approx 10\ \rm{Pa\, s}$ \cite{Rubie03}. To do so, we choose $\phi_c=60.002\%$ and which yields $\eta(T=T_l^{\rm{Sil}})=14\ \rm{Pa\, s}$. The rheological transition described here is plotted in Figure \ref{fig:RT} assuming batch melting, so that the temperature is a direct measure of the solid content $\phi$ given by (\ref{eq:phi}). Figure \ref{fig:RT} shows that the viscosity decreases at the rheological transition by at least 7 orders of magnitude. Beyond this transition, the proto-mantle behaves like a liquid suspension called ``magma ocean", according to \citeA{Taylor92}.\\
\indent We can now complete the approach of \citeA{Kaminski20} by giving both the time at which solid-state convection occurs and the onset time of the magma ocean, i.e.: the time at which the temperature in the planetesimal reaches $T_{\rm{RT}}=1390\rm{^oC}$. Results displayed in Figure \ref{fig:SSC_MO} show three types of evolutions. (i) For small planetesimals ($R<30\ \rm{km}$), the surface over volume ratio is high enough to prevent the onset of solid state convection, and all the more a magma ocean episode. In this purely conductive regime, the planetesimals remain undifferentiated and relatively cold. (ii) For planetesimals with radius $30\ \rm{km}<R<400\ \rm{km}$, the temperature is high enough to sufficiently melt the planetesimal and to trigger convection. However, as they are small bodies, the onset time for solid-state convection is long, and the planetesimals reached the rheological transition before solid-state convection has started. Hence, these planetesimals experienced an initial conductive stage for $\approx 0.25-0.4\ \rm{Myr}$, before a direct transition to a magma ocean. (iii) Larger planetesimals ($R>400\ \rm{km}$) undergo first a conductive regime, followed by a short solid-state convective regime, before finally reaching the rheological transition. As shown in Figure \ref{fig:SSC_MO}, the onset of convection happens early ($0.25-0.4\ \rm{Myr}$) and the larger the body, the earlier the onset.\\
\indent We now have constrained the planetesimal size required to produce a magma ocean. Our model shows that even small planetesimals can experience a magma ocean episode. Once the rheological transition is reached, the system behaves like a liquid suspension of crystals. These crystals can then sediment or float accordingly to their buoyancy. In the following, we study crystal segregation from a convective magma ocean.

\begin{table}
	\centering
	\begin{tabular}{ c c c c  c }
	\hline
	Parameters	 				& Symbols				& Values				& Units					& Ref.\\
	\hline
	Iron content 					& $x_{\rm{Fe}}$			& 18					& wt$\%$					& (4) \\
	Silicate content 				& $x_{\rm{Sil}}=1-x_{\rm{Fe}}$	& 82					& wt$\%$					& (4) \\
	\hline
	Chondritic abundance ($^{60}Fe$)	& $F_{\rm{Fe}}$			& $2.14.10^{24}$		& $\rm{kg^{-1}}$			& (1)\\
	Chondritic abundance ($^{26}Al$)	& $F_{\rm{Al}}$				& $2.62.10^{23}$		& $\rm{kg^{-1}}$			& (1)\\
	Decay energy per atom ($^{60}Fe$)	& $\mathcal{E}_{\rm{Fe}}$		& $4.87.10^{-13}$		& $\rm{J}$					&(1)\\
	Decay energy per atom ($^{26}Al$)	& $\mathcal{E}_{\rm{Al}}$		& $4.42.10^{-13}$		& $\rm{J}$					&(1)\\
	Initial ratio ($^{60}Fe$)			& $[\rm{^{60}Fe/^{56}Fe}]_0$	& $10^{-8}$			& -						& (2)\\
	Initial ratio ($^{60}Fe$)			& $[\rm{^{26}Al/^{27}Al}]_0$	& $5.10^{-5}$			& -						& (2)\\
	Half life ($^{60}Fe$)				& $t_{1/2}^{Fe}$			& $2.6$				& Myr					&(1)\\
	Half life ($^{26}Al$)				& $t_{1/2}^{Al}$				& $0.717$				& Myr					&(1)\\
	\hline

	Iron density					& $\rho_{\rm{Fe}}$			& $7800$				& $\rm{kg\, m^{-3}}$			& (4)\\
	Silicate density					& $\rho_{\rm{Sil}}$			& $3200$				& $\rm{kg\, m^{-3}}$			& (4)\\
	Crystal/melt drop of density		& $\Delta \rho$				& $100$				& $\rm{kg\ m^{-3}}$			& -\\
	Iron thermal conductivity			& $\lambda_{\rm{Fe}}$		& $50$				& $\rm{W\, m^{-1}\, K^{-1}}$	& (4)\\
	Silicate thermal conductivity		& $\lambda_{\rm{Sil}}$		& $3$				& $\rm{W\, m^{-1}\, K^{-1}}$	& (4)\\
	Thermal expansion (Fe)			& $\alpha_{\rm{Fe}}$			& $7.7.10^{-5}$			& $\rm{K^{-1}}$				& (1)\\
	Thermal expansion (Silicate)		& $\alpha_{\rm{Sil}}$			& $2.10^{-5}$			& $\rm{K^{-1}}$				& (1)\\
	Specific heat (Silicate)			& $c_{p,\rm{Sil}}$			& $1168$				& $\rm{J\, K^{-1}\, kg^{-1}}$	& (3)\\	
	Specific heat (Fe)				& $c_{p,\rm{Fe}}$			& $622$				& $\rm{J\, K^{-1}\, kg^{-1}}$	& (3)\\
	Latent heat ($\rm{Fe-FeS}$)		& $L_{\rm{Fe}}$			& $250$				& $\rm{J\, kg^{-1}}$			& (4)\\
	Latent heat (Silicate)				& $L_{\rm{Sil}}$			& $500$				& $\rm{J\, kg^{-1}}$			& (4)\\
	Liquidus ($\rm{Fe-FeS}$)			& $T^{\rm{Fe}}_{l}$			& $1615$				& $\rm{^oC}$				& (4)\\
	Solidus ($\rm{Fe-FeS}$)			& $T^{\rm{Fe}}_{s}$			& $990$				& $\rm{^oC}$				& (4)\\
	Liquidus (Silicate)				& $T^{\rm{Sil}}_{l}$			& $1800$				& $\rm{^oC}$				& (4)\\
	Solidus (Silicate)				& $T^{\rm{Sil}}_{s}$			& $1150$				& $\rm{^oC}$				& (4)\\
	Solidus light mineral 				& $T^{\rm{L}}_s$			& $1150$				& $\rm{^oC}$				& (6)\\
	Solidus heavy mineral			& $T^{\rm{H}}_s$			& $1280$				& $\rm{^oC}$				& (6)\\
	Liquidus light mineral			& $T_l^{\rm{L}}$			& $1605$				& $\rm{^oC}$				& (6)\\
	Liquidus heavy mineral 			& $T_l^{\rm{H}}$			& $1800$				& $\rm{^oC}$				& (6)\\
	Activation energy ($\rm{Sil_{(l)}}$)	& $E_{a,\rm{Sil}}$			& $250$				& $\rm{kJ\, mol^{-1}}$		& (4)\\
	Activation energy ($\rm{Fe_{(l)}}$)	& $E_{a,\rm{Fe}}$			& $13$				& $\rm{kJ\, mol^{-1}}$		& (4)\\
	Viscosity (Iron melt at $T^{\rm{Fe}}_{l}$)& $\eta_{0}$				& $1$				& $\rm{Pa\, s}$				& (3)\\
	Melt fraction factor				& $\sigma_{Sil}$			& $21$				& - 						& (4)\\
	Crystals radius					& $r$					& $3.10^{-3}$			& $\rm{m}$				& (5)\\
	\hline
	\end{tabular}
	\caption{Parameters used in the thermal model. References are the following: \citeA{Neumann14} (1), \citeA{Tang12} (2), \citeA{Neumann12} (3), \citeA{Kaminski20} (4), \citeA{Solomatov00} (5), adapted and simplified from \citeA{Righter97} (6).}
	\label{tab:paramsMGP}
\end{table}

%
%
		
	\subsection{Magma ocean, core formation and proto-crust}
	\indent Once the rheological transition is reached, two differentiations may occur: (i) the metal-silicate segregation and (ii) the crystal/melt segregation in the magma ocean.\\
	\indent We assume here that core-mantle differentiation occurs in the route to the rheological transition. Here we assume that it is total and occurs instantaneously at the rheological transition as the fluid becomes drastically less viscous \cite{Hoink06}. All the iron gets into the core and mass conservation gives the following relationship between the ratio $R_c/R$ and densities:
\begin{linenomath*}
	\begin{equation}
		f_c=\frac{R_c}{R}=\left( \frac{\rho}{\rho_{\rm{Fe}}}\, x_{\rm{Fe}} \right)^{1/3},
	\end{equation}
\end{linenomath*}		
	which yields $f_c=0.44$ with values summarized in Table \ref{tab:paramsMGP}. All $^{26}\rm{Al}$ is partitioned into the silicate magma ocean, and all the $^{60}\rm{Fe}$ is stored in the core. \\
	\indent Before the rheological transition, solid-state convection takes place under a stagnant lid whose thickness is about $1\, \rm{km}$ (\ref{App:SLC}). By definition, this conductive lid is disrupted neither by convection nor by partial melting. The conductive lid stands thus as an undifferentiated layer preserved at the surface of the magma ocean, a classical hypothesis in the literature \cite{Haack90, ElkinsTanton11, ElkinsTanton11b, Mandler13}. \\
	\indent Below the relic of the stagnant lid, the magma ocean is a convective liquid suspension from which crystals can segregate to form a floating crust and/or a basal cumulate. As highlighted in the companion paper, the formation of a flotation crust in particular can further control the thermal evolution of the magma ocean.

%
%

\section{Beyond the rheological transition: sedimentation/flotation of crystals}

%
%
		
	\subsection{Energy budgets}
	\indent Energy budgets in the magma ocean and the core lead to the following evolution equations of the average temperature of the magma ocean $T_{\rm{MO}}$ and the core $T_c$:
\begin{linenomath*}
	\begin{eqnarray}
		\rho_{\rm{Sil}}c_{p,\rm{Sil}}\frac{\rm{d}T_{\rm{MO}}}{\rm{d}t} &=& H_{\rm{MO}}+Q_{\rm{CMB}}\,  \frac{S_c}{V_m} - Q_{s}\, \frac{S_{m}}{V_{m}}, \label{eq:BudgetMO} \\
		\rho_{\rm{Fe}}c_{p,\rm{Fe}}\frac{\rm{d}T_c}{\rm{d}t} &=& H_{c}- Q_{\rm{CMB}}\, \frac{S_c}{V_c}, \label{eq:BudgetCore}
	\end{eqnarray}
\end{linenomath*}		
	where $Q_{s}$ is the surface heat flux, $Q_{\rm{CMB}}$ is the heat flux at the core mantle boundary (CMB), $H_{i}$ is the volumetric rate of internal heating in the layer $i$, $S_m=4\pi R^2(1-\delta_{\rm{L}}/R)^2$ is the surface area of the magma ocean, $V_m=4/3\pi R^3 [(1-\delta_{\rm{L}}/R)^3-(f_c+\delta_{\rm{H}}/R)^3]$ is the volume of the magma ocean, $S_c=4\pi R^2 f_c^2$ is the surface of the core,  $V_c=4/3\pi R^3 f_c^3$ is its volume, $\delta_{\rm{L}}$ is the thickness of the flotation crust and $\delta_{\rm{H}}$ is the thickness of the cumulate at the CMB. Note that we consider the flux positively upward.\\
	\indent The internal heat sources are twofold: (i) one part is due to energy supplied by radioactive decay, (ii) the other part is due to latent heat released during crystallization. Assuming $\rm{^{60}\rm{Fe}}$ to partitioned into the core and $\rm{^{26}\rm{Al}}$ into the silicate magma ocean, the corresponding heating rates are calculated by multiplying (\ref{eq:Fe}) and (\ref{eq:Al}) by a concentration factor:
\begin{linenomath*}
	\begin{eqnarray}	
	 H_{\rm{Fe}}		&=&		H_{0,\rm{Fe}}\, f_c^{-3}, \label{eq:Hstar}\\
	 H_{\rm{Al}}		&=&		H_{0,\rm{Al}}\, \left[ \left(1-\frac{\delta_{\rm{L}}}{R}\right)^3 -\left(f_c+\frac{\delta_{\rm{H}}}{R}\right)^3 \right]^{-1}, \label{eq:Hstar}
	\end{eqnarray}
\end{linenomath*}	 
whereas the heat released by latent heat in each reservoir is:
\begin{linenomath*}
	\begin{eqnarray}
	H_{LH,\rm{Fe}}		&=&		L_{\rm{Fe}}\, x_{\rm{Fe}}\, \frac{\rho}{\rho_{\rm{Fe}}}\, \frac{\rm{d}T_{\rm{MO}}}{\rm{d}t}\, f_c^{-3},\\	
	 H_{LH,\rm{Sil}}		&=&		L_{\rm{Sil}}\, x_{\rm{Sil}}\, \frac{\rho}{\rho_{\rm{Sil}}}\, \frac{\rm{d}T_{\rm{MO}}}{\rm{d}t}\, \left[ \left(1-\frac{\delta_{\rm{L}}}{R}\right)^3 -\left(f_c+\frac{\delta_{\rm{H}}}{R}\right)^3 \right]^{-1}.
	\end{eqnarray}
\end{linenomath*}	 
We get the internal heating rate for the magma ocean and the core, respectively: $H_{\rm{MO}}=H_{\rm{Al}}+H_{LH,\rm{Sil}}$ and $H_c=H_{\rm{Fe}}+H_{LH,\rm{Fe}}$\\
	\indent The heat fluxes are expressed using the scaling laws for convection \cite{Limare19,Kaminski20,Limare21}. The surface heat flux is given by:
\begin{linenomath*}
	\begin{equation}
		Q_{s}=\lambda_{\rm{Sil}}\, \left(\frac{\alpha_{\rm{Sil}}\rho_{\rm{Sil}}g}{\kappa_{\rm{Sil}}\eta_{\rm{Sil}}}\right)^{1/3}\, \left(\frac{T_{\rm{MO}}-T_{\rm{lid}}}{C_T}\right)^{4/3},
	\end{equation}
\end{linenomath*}
where $g=4/3\pi \mathcal{G} \rho R$ the surface gravity, with $\mathcal{G}$ the gravitational constant, and $T_{\rm{lid}}$ is the temperature at the base of the crust. The temperature $T_{\rm{lid}}$ is determined assuming that heat is transported by conduction in the crust which yields:
\begin{linenomath*}
	\begin{equation}
		Q_s=\lambda_{\rm{Sil}}\, \frac{T_{\rm{lid}}-T_s}{\delta_L(t)}\, \frac{R(t)}{R(t)-\delta_L(t)},
	\end{equation}
\end{linenomath*}
\indent For the CMB heat fluxes, two cases has to be considered. If the core is colder than the magma ocean,  a TBL exists at the base of magma ocean and the heat flux $Q_{\rm{CMB}}$ is given by:
\begin{linenomath*}
	\begin{equation}
		Q_{\rm{CMB}}=\left(\frac{\alpha_{\rm{Sil}}\rho_{\rm{Sil}}g_c}{\kappa_{\rm{Sil}}\eta}\right)^{1/3}\, \left(\frac{T_{\rm{MO}}-T_{b,\rm{MO}}}{C_T}\right)^{4/3},
	\end{equation}
\end{linenomath*}
where $g_c=4/3\pi \mathcal{G} \rho_{\rm{Fe}} R_c$ is the gravity at the top of the core and $T_{b,\rm{MO}}$ is the temperature at the top of the cumulate. The temperature $T_{b,\rm{MO}}$ is determined assuming conductive heat transfer in the cumulate and an isothermal core at temperature $T_c$ which yields:
\begin{linenomath*}
	\begin{equation}
		Q_{\rm{CMB}}=\lambda_{\rm{Sil}}\, \frac{T_{b,\rm{MO}}-T_{\rm{CMB}}}{\delta_H(t)}\frac{R_c(t)+\delta_H(t)}{R_c(t)},
	\end{equation}	
\end{linenomath*}
If the magma ocean is colder than the core, we consider a TBL at the top of the core, beyond the cumulate whose basal temperature is $T_{b,\rm{CMB}}$. In this case, the CMB heat flux is given by the scaling law for convection in the core:
\begin{linenomath*}
	\begin{equation}
		Q_{\rm{CMB}} = \left(\frac{\alpha_{\rm{Fe}}\rho_{\rm{Fe}}g_c}{\kappa_{\rm{Fe}}\eta_{\rm{Fe}}}\right)^{1/3}\, \left(\frac{T_{c}-T_{b,\rm{CMB}}}{C_T}\right)^{4/3},
	\end{equation}
\end{linenomath*}
and $T_{b,\rm{CMB}}$ is deduced from the conductive heat transfer through the cumulate, assuming a top temperature equal to the magma ocean one which yields:
	\begin{equation}
		Q_{\rm{CMB}}=\lambda_{\rm{Sil}}\, \frac{T_{b,\rm{CMB}}-T_{\rm{MO}}}{\delta_H(t)}\frac{R_c(t)+\delta_H(t)}{R_c(t)}.
	\end{equation}	
\indent The resolution of this set of equations relies on the knowledge of two parameters: the thickness of the flotation crust $\delta_L$ and the thickness of the basal cumulate $\delta_H$. These two thicknesses can be obtained using the scaling laws established in the companion paper \cite{Sturtz21b}.

%
%
 
 	\subsection{Mass conservation and deposits evolution}

\indent The magma ocean contains a total volume of solid crystals $\phi_0$ (given by (\ref{eq:phi})). The possibility to sustain crystals in suspension and/or to form deposits are described by \citeA{Sturtz21}. The criteria is based on the Shields number that compare the strength of convective stress and the buoyancy of particles, and that is defined as follows:
\begin{linenomath*}
	\begin{equation}
		\zeta_{\rm{MO}}	=		\frac{\eta_{\rm{Sil}} \kappa_{\rm{Sil}}}{\Delta \rho g r h_{MO}^2}\, (Ra_{H,\rm{Sil}}^*)^{3/8}, \label{eq:zeta}
	\end{equation}
\end{linenomath*}
with $Ra_H^*$ the modified Rayleigh-Roberts number for time-varying system:
\begin{linenomath*}
	\begin{equation}
		Ra_{H,\rm{MO}}^*	=		\frac{\alpha_{\rm{Sil}}\rho_{\rm{Sil}}g H^* h_{\rm{MO}}^5}{\kappa_{\rm{Sil}} \eta_{\rm{Sil}} \lambda_{\rm{Sil}}},
	\end{equation}
\end{linenomath*}
with $H^*$ the effective heating rate given by:
\begin{linenomath*}
	\begin{equation}
		H^*	=	H_{\rm{Sil}}-\rho_{\rm{Sil}}c_{p,\rm{Sil}}\frac{\rm{d}T_{\rm{MO}}}{\rm{d}t},
	\end{equation}
\end{linenomath*}
and $h_{MO}$ the thickness of the magma ocean:		
\begin{linenomath*}
	\begin{equation}
		h_{\rm{MO}}		=		(1-f_c)\, R-\delta_{\rm{L}}(t)-\delta_{\rm{H}}(t).
	\end{equation}
\end{linenomath*}
If $\zeta_{\rm{MO}}$ is larger than the critical value $\zeta_c=0.29\pm0.17$, the convection vigor is large enough to destabilize any cumulate previously formed and to prevent suspended crystals from settling and/or floating. But if $\zeta_{\rm{MO}}<\zeta_c$, the buoyancy of particles is large enough for them to segregate from the liquid suspension and to form a flotation crust and/or a basal cumulate. The sharp variation of viscosity that occurs at the rheological transition makes $\eta$ the key parameter for the segregation of crystals as $\zeta_{\rm{MO}}\sim \eta^{5/8}$. As illustrated in Figure \ref{fig:RT}, segregation of particles is possible in the magma ocean episode, whereas solid-state convection corresponds to a regime where pre-existing cumulates are unstable and tend to be eroded by the convection. This approach highlights that the only way for a planetesimal to form differentiated crystal layer is to undergo a magma ocean episode.  During a magma ocean episode, deposits previously formed are preserved as convective shear stress is too low to erode them. Even if convection is turbulent (very high Rayleigh-Roberts number), the viscosity is so small that the Shields number will never reach the critical value, hence crystal layers will be always stable, and potentially growing.\\
\indent The evolution of both layers thickness $\delta_i$ is given by the deposition law described in the companion paper \cite{Sturtz21b}:
\begin{linenomath*}
	\begin{equation}
		\frac{\rm{d}\delta_i}{\rm{d}t}=c_d v_s  \frac{ \overline\phi_{i,\rm{sus}}}{\phi_{\rm{RLP}}}. \label{eq:CumulateEvol}
	\end{equation}
\end{linenomath*}
 where $v_s$ is the settling velocity, $c_d$ is a constant, $\phi_{i,\rm{sus}}$ the volume fraction of crystal $i$ in suspension that contributes to the thickening of the corresponding deposit, and $\phi_{\rm{RLP}}$ is the packing inside the deposit. The volume fraction of crystals that remains in suspension is given by mass conservation as explained in \ref{App:MassCons}.\\
\indent Experimentally, \citeA{Sturtz21b} pointed out that the sedimentation velocity $v_s$ in presence of convection scales with the Stokes velocity of particles, with $c_d=0.24\pm0.14$ close to the 2/9 coefficient of the Stokes velocity. This reasoning is consistent with a diluted suspension. However, it does not take into account particle-particle interactions, that are significant for dense suspensions. Therefore, for geophysical application, we re-define the settling velocity $v_s$ as the difference between the convective velocity $\mathbf{u}_f$ and the velocity of particles $\mathbf{u}_p$. At large crystal content, this velocity tends to the rate at which magma is drained from a slurry, and thus, we model the drop of velocity by a modified Darcy law \cite{Neumann12,Neumann14}:
\begin{linenomath*}
	\begin{eqnarray}
		v_s=||\mathbf{u}_p-\mathbf{u}_f||&=&\frac{K_{\phi}\Delta \rho g}{(1-\phi)\eta_l}, \label{eq:vs1}\\
		K_{\phi}&=&\frac{r^2(1-\phi)^n}{\tau}, \label{eq:vs2}
	\end{eqnarray}
\end{linenomath*}
where $K_{\phi}$ is the permeability, $\eta_l$ is the dynamic viscosity of the liquid magma, and $(\tau,\, n)=(1600,2)$ according to \citeA{Neumann12}. This law is applicable for all values of $\phi$, and for $\phi=0$, $v_s$is equal to the Stokes velocity. As the crust is formed by flotation of light crystals, the temperature at its base can not be larger than the liquidus $T_l^{\rm{L}}$. If this should happen, we would decreases the thickness of the crust to ensure that the basal temperature is at the liquidus.\\
\indent  If the Shields number becomes super-critical, i.e.: $\zeta_{\rm{MO}}>\zeta_c$, then the deposits are eroded according to the erosion law described in the companion paper \cite{Sturtz21b}:
\begin{linenomath*}
	\begin{equation}
		\frac{\rm{d} \delta}{\rm{d}t} =-c_e\, \frac{\kappa_{\rm{Sil}} r}{h_{\rm{MO}}^2}\, Ra_{H,\rm{MO}}^{*,1/2}\, (\zeta_{\rm{MO}}-\zeta_c), \label{eq:LidEvolution}
	\end{equation}
\end{linenomath*}
with $c_e=1.0\pm0.8$. 

%
%

\section{Magma ocean dynamics - crystal/melt segregation}

%
%

	\subsection{Reference evolution}
\begin{figure}
	\centering
	\includegraphics[width=0.6\textwidth]{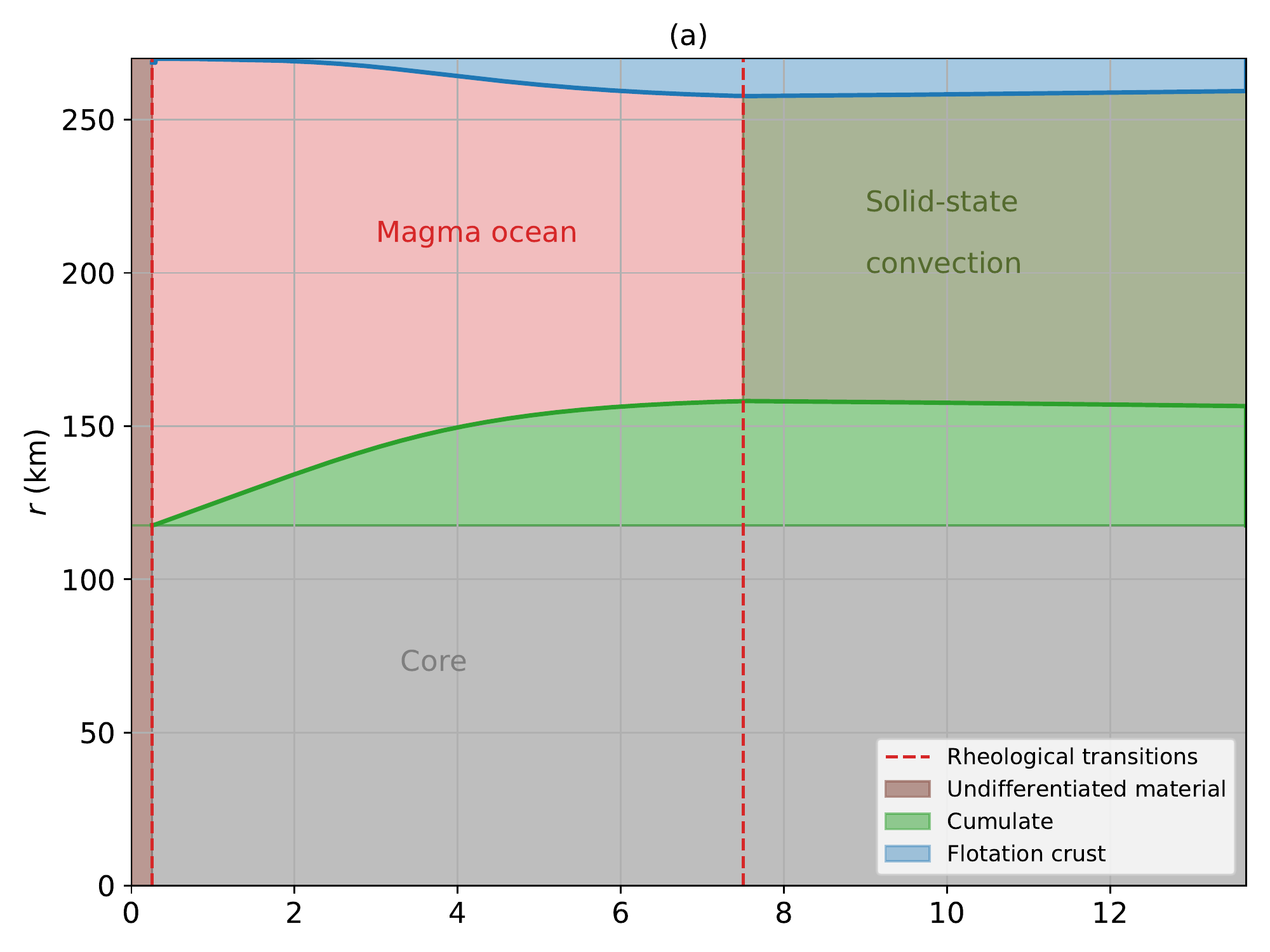}
	\includegraphics[width=0.6\textwidth]{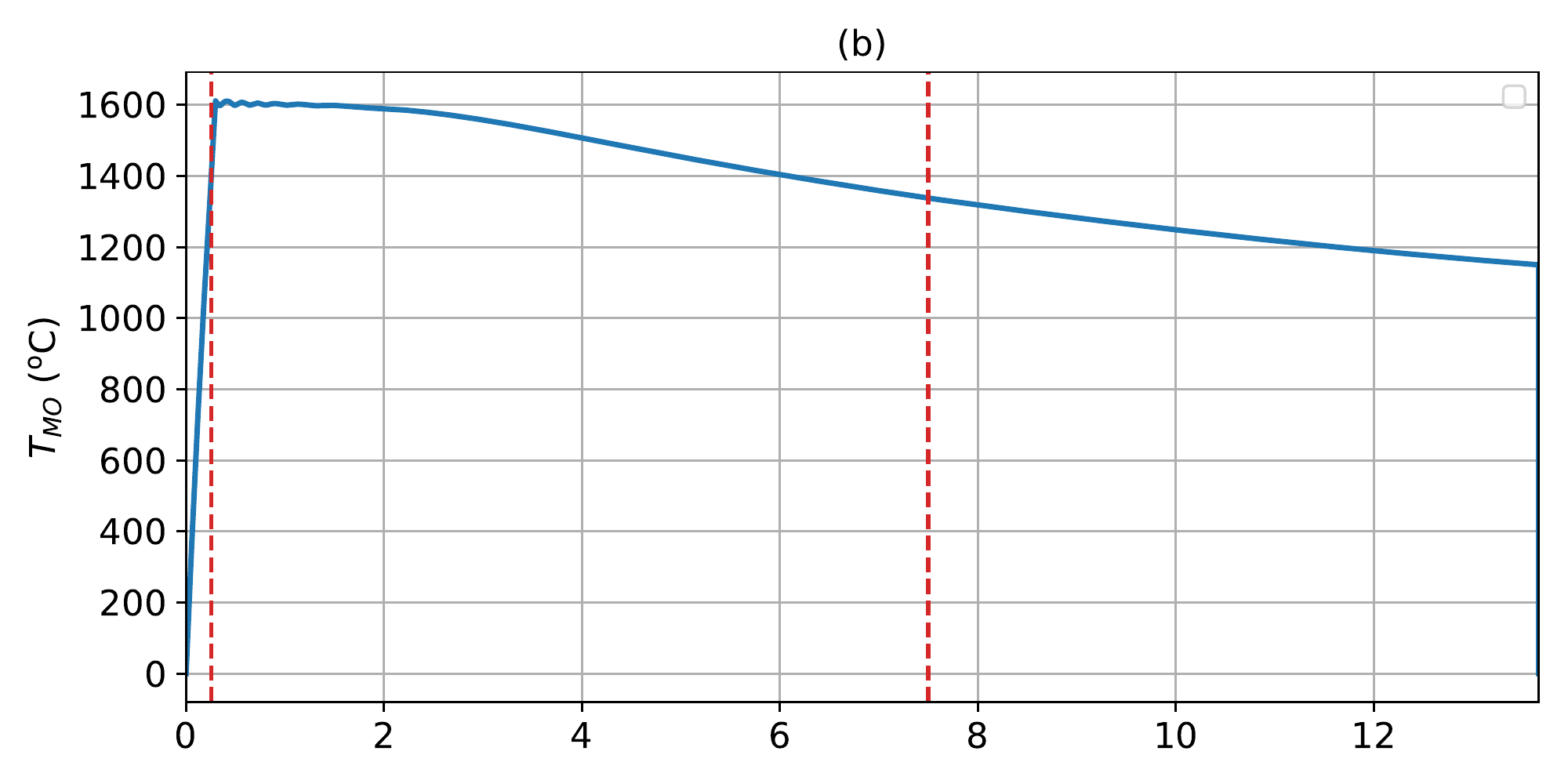}
	\includegraphics[width=0.6\textwidth]{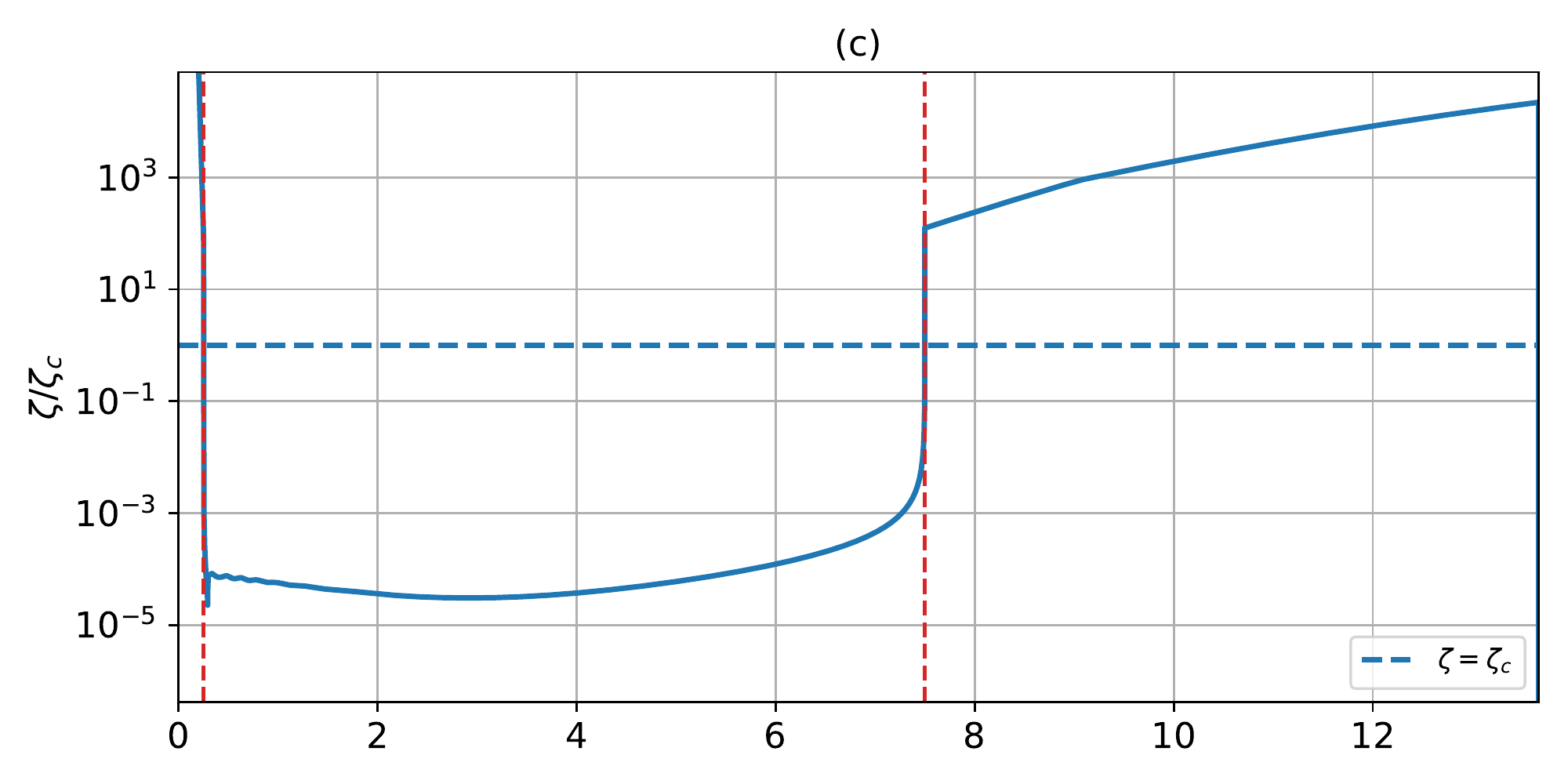}
	\includegraphics[width=0.6\textwidth]{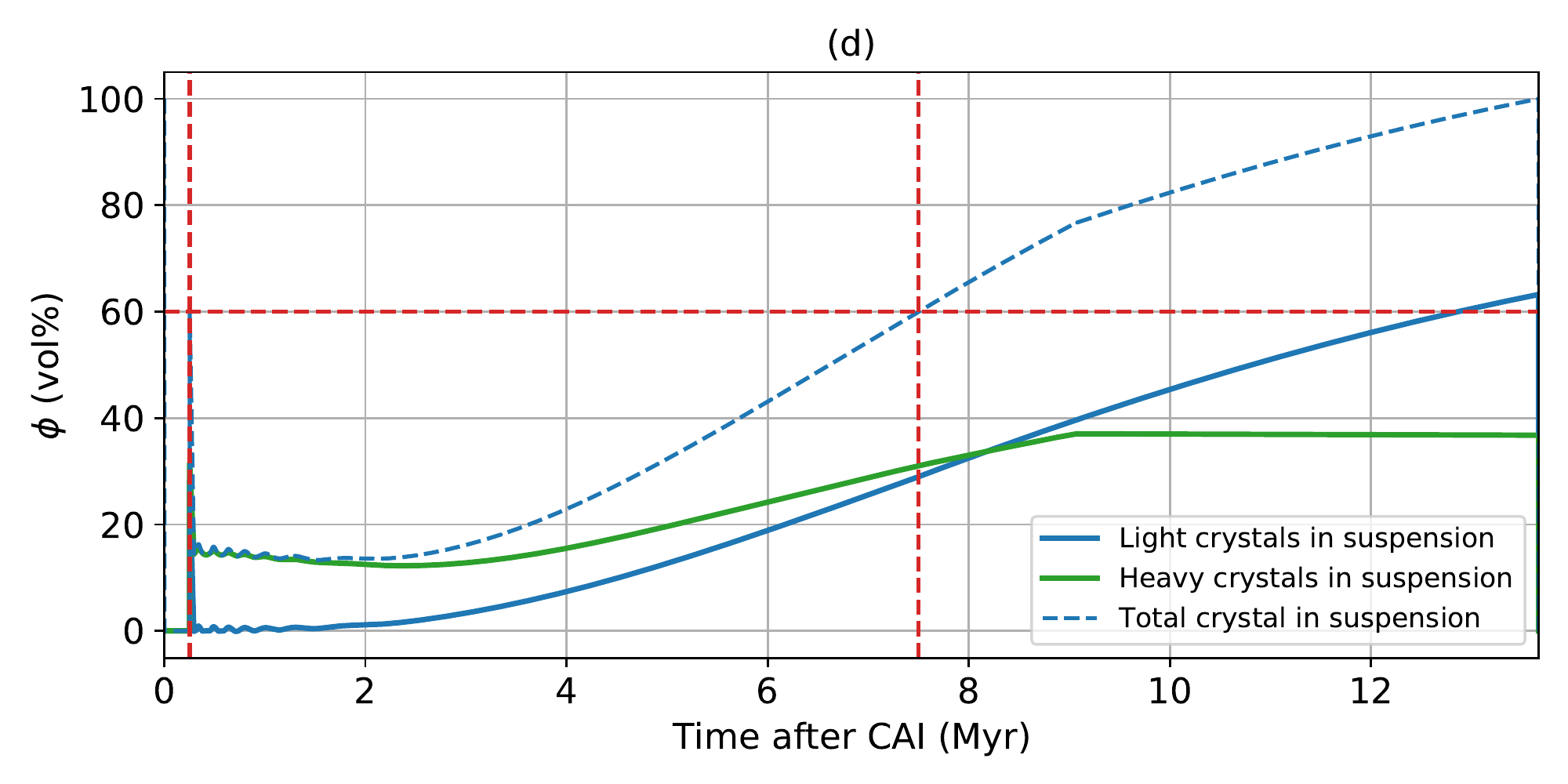}
	\caption{(a) Evolution of the radial structure of a planetesimal of 270 radius instantaneously accreted at $t_0=0$ Myr, (b) bulk temperature, (c) Shields number, (d) volume fraction of crystals in suspension.}
	\label{fig:R270}
\end{figure}
\begin{figure}
	\centering
	\includegraphics[width=\textwidth]{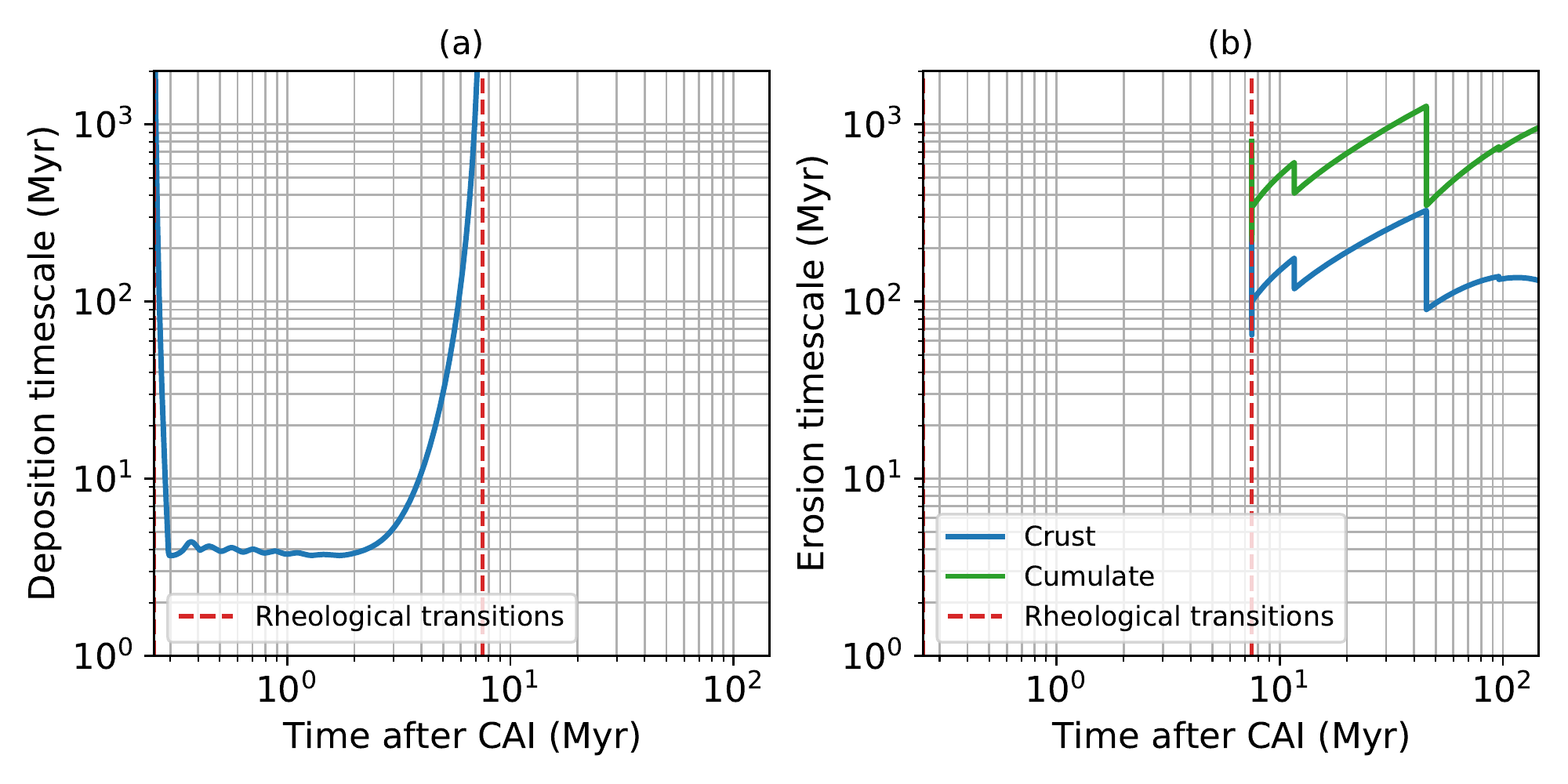}
	\caption{Evolution of the deposition time scale  (a) for flotation crust and the basal deposit (same time scale for the two layers), and evolution of the erosion timescale (b) for the crust (blue line) and for the cumulate (green line) for a 270 km radius planetesimal.}
	\label{fig:taudeptauero}
\end{figure}

	\indent We first apply the model presented above to a planetesimal with radius $R=270\ \rm{km}$. Figure \ref{fig:R270} (a) shows the evolution of the internal structure. After a short episode of conductive regime spanning from $0$ to $0.3$ Myr, the rheological transition is reached and the magma ocean episode begins with $60\, \rm{vol}\%$ of crystals in suspension (Figure \ref{fig:R270} (d)). During all that stage, the Shields number in the magma ocean is far below the critical value (Figure \ref{fig:R270} (c)), and crystals can settle or float according to their buoyancy. The cumulate forms right after the rheological transition, whereas light crystals float only when the temperature becomes smaller than $T_{l}^{\rm{L}}$ at $2\, \rm{Myr}$, and can then form a flotation crust. The rate of temperature decrease is large enough for the nucleation rate to remain always larger than the rate of segregation of crystals from the suspension, hence the concentration of crystals in suspension increases. At 7.5 Myr, the rheological transition is reached again. At this time, the dynamics transitions from that of a liquid to that of a solid-state convection, leading to a Shields number much greater than the critical value, which prevents further sedimetation/flotation and triggers erosion. We later stop the simulations when the mantle temperature reaches the solidus temperature of silicate, at 14 Myr, when the two-phase framework does not apply anymore.\\
	\indent We describe now more quantitatively the erosion/deposition dynamics. To quantify the deposition time scale, we refer to the deposition law (\ref{eq:CumulateEvol}). Using the settling velocity $v_s$ that has been chosen, the timescale for deposition can be approximated by $\tau_{d}\approx h_{\rm{MO}}/v_s$ i.e.:
\begin{linenomath*}
	\begin{equation}
		\tau_d=\frac{h_{\rm{MO}}}{R}\frac{3 \eta_l \tau}{4\pi \mathcal{G} \rho \Delta \rho r^2 (1-\phi)^{n-1}}.\label{eq:taud}
	\end{equation}
\end{linenomath*}
This timescale is calculated in the case exposed before for $R=270\ \rm{km}$, and the result is shown in Figure \ref{fig:taudeptauero} (a). At the beginning of the magma ocean episode, the timescale for deposition is smaller (3 Myr) than the magma ocean life time (equal to 7 Myr in this case), allowing deposition to occur. As the magma ocean cools down, it bears more and more crystals that increase its viscosity and thus the deposition timescale. Consequently, the deposition rate is reduced, as displayed in Figure \ref{fig:taudeptauero} (a), until is stopped at the rheological transition.\\
\indent Once the magma ocean has ended, the planetesimal experiences again solid-state convection and thus deposition stops and erosion starts as the Shields number is very high (Figure \ref{fig:R270} (c)) and the rate of erosion is:
\begin{linenomath*}
	\begin{equation}
		\frac{\rm{d}\delta}{\rm{d}t} \approx -c_e\frac{\kappa_{\rm{Sil}} r}{ h_{\rm{MO}}^2}\, Ra_{H,\rm{MO}}^{*,1/2}\, \zeta_{\rm{MO}}. \label{eq:dddt}	
	\end{equation}
\end{linenomath*}
Combining (\ref{eq:dddt}) with (\ref{eq:zeta}), a natural erosion velocity $V_E$ appears:
\begin{linenomath*}
	\begin{equation}
		V_E=\frac{3}{4\pi }\, \frac{\kappa_{\rm{Sil}}^2 \eta}{R h_{\rm{MO}}^4\mathcal{G}\rho \Delta \rho}\, Ra_{H,\rm{MO}}^{*,7/8},
	\end{equation}
\end{linenomath*}
and the timescale for erosion is $\delta_{\rm{L}}/V_E$ for the crust and $\delta_{\rm{H}}/V_E$ for the cumulate. Interestingly, this erosion velocity is independent of the crystal radius $r$, even though the model relies on the erosion of a granular media. These timescales are displayed in Figure \ref{fig:taudeptauero} (b), it shows that it would take over 100 Myr to erode the crust and around 1 Gyr to recycle the cumulate. The discontinuities in the erosion timescales are due to the slope of the viscosity that changes when the soliduses are reached. These times are larger than the whole life-time of the solid-state convection regime, which is about 150 Myr for a 270 km planetesimal (based on the hypothesis that convection stops when $Ra_H^*$ becomes smaller than its critical value). As a result, even though deposits are unstable according to the Shields' criteria, they are kinetically preserved as their erosion timescales are longer than the duration solid-state convection episode.

%
%

	\subsection{Layered planetary bodies - from homogeneous mantle to "onion shell" interior}
	
	\begin{figure}
		\centering
		\includegraphics[width=0.7\textwidth]{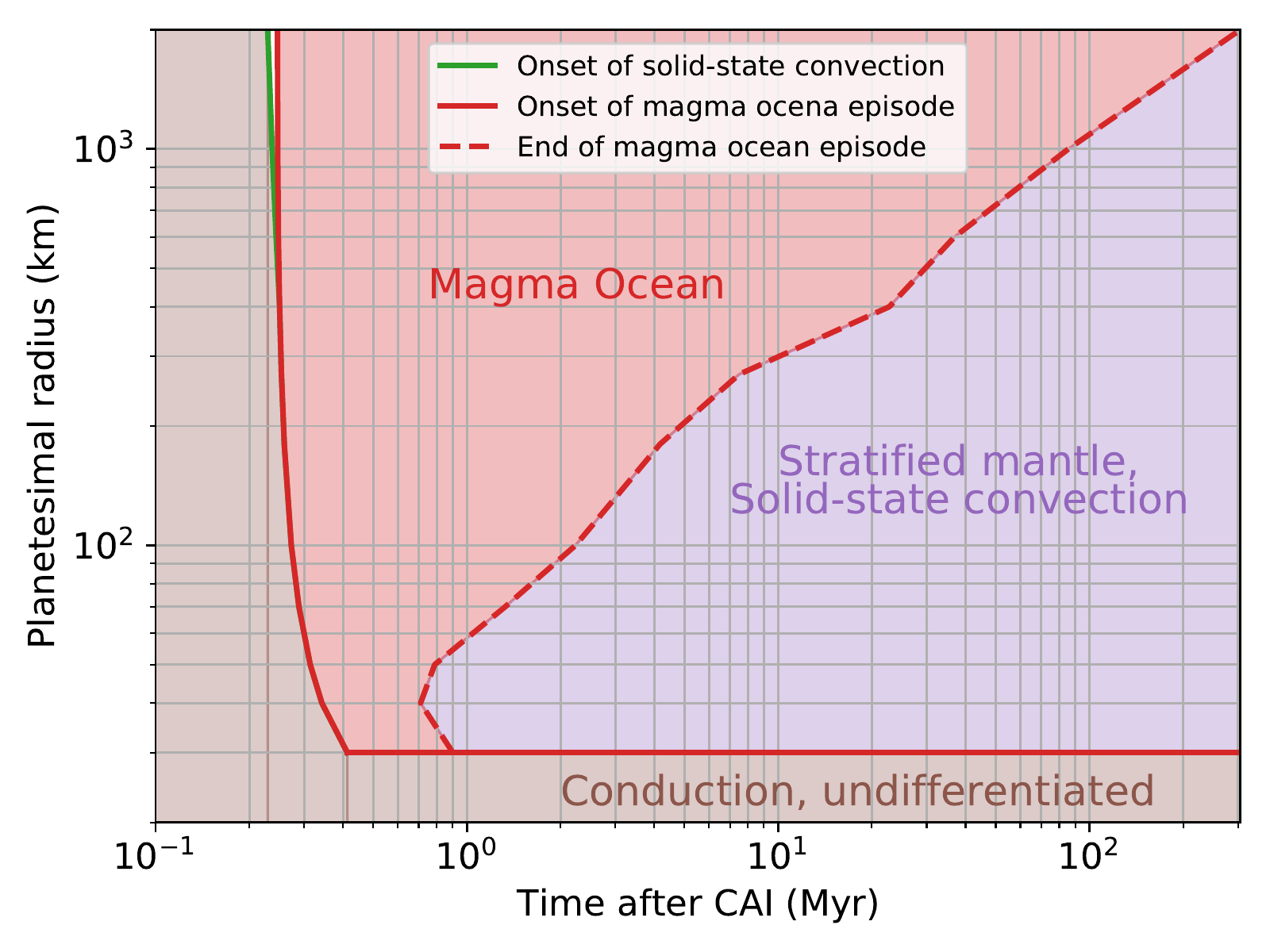}
		\caption{Thermal history of instantaneously accreted planetesimals as a function of their radius.}
		\label{fig:EvolDiagram}
	\end{figure}
	
	\begin{figure}
		\centering
		\includegraphics[width=0.7\textwidth]{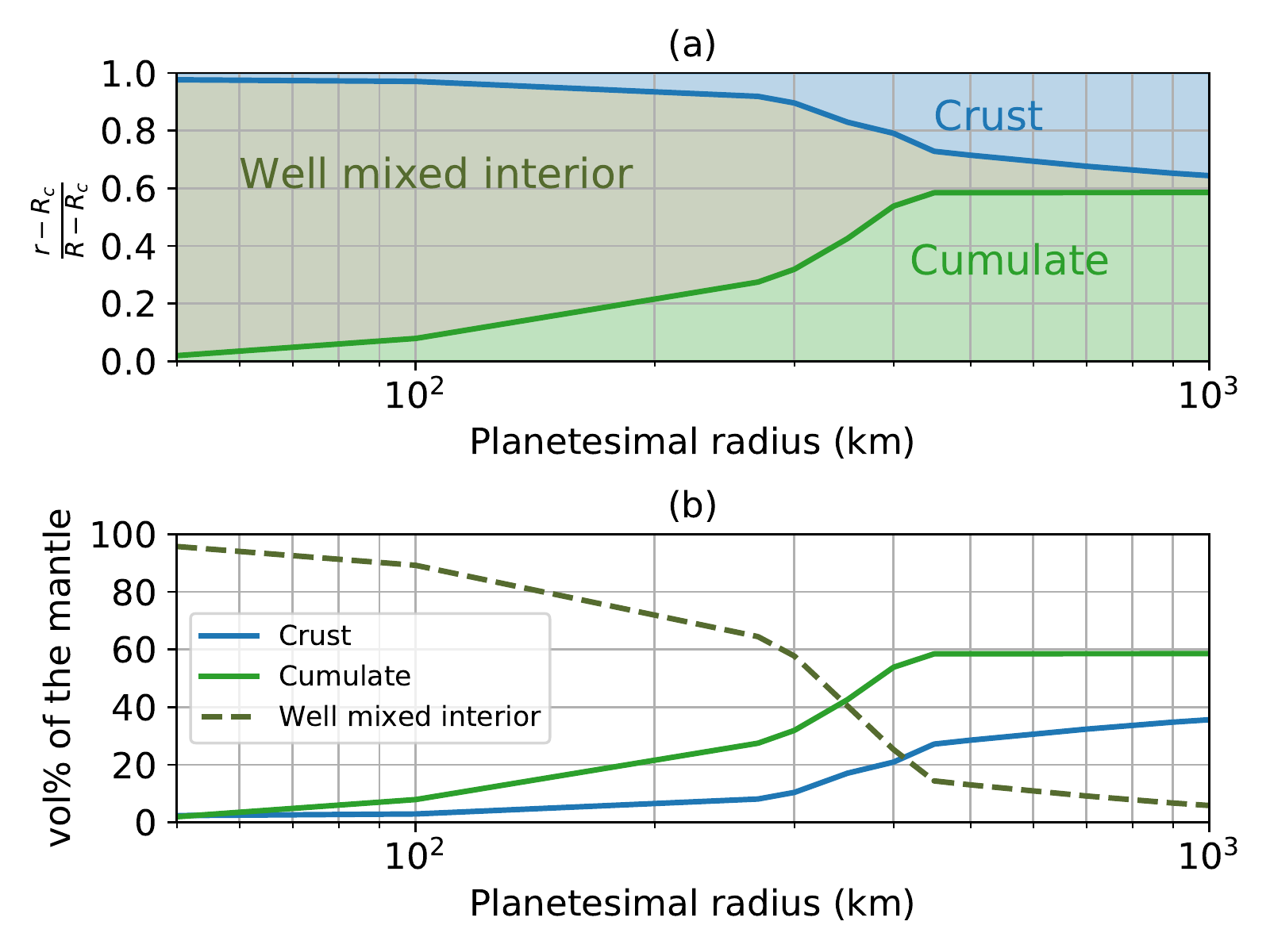}
		\caption{Dimensionless thicknesses of the crust, of the cumulate and of the well mixed interior at the end of the magma ocean episode, when the rheological transition is reached again (a) as a function of the planetesimal radius. (b) The corresponding fraction of the mantle occupied by each layer at the end of the magma ocean episode as a function of the planetesimal radius.}
		\label{fig:solidification}
	\end{figure}
	
\indent We now consider the evolution of the internal structure of the planetesimal mantle as a function of the planetesimal size, all other things being equal. We perform numerical simulations for planetesimals with radius $R\in [10,\, 2000]\, \rm{km}$, leading to the complete planetesimal evolution diagram of Figure \ref{fig:EvolDiagram}. The three scenarios described in section \ref{subsec:RT} remain relevant, but we further add the duration of the magma ocean lifetime. For $R>30\, \rm{km}$, the magma ocean episode lifetime increases with the planetesimal size, from tens Myr to 1 Gyr. This is due to the energy released by the radiogenic elements during the first few Myr that is stored in the magma ocean and is only partially removed because of the thermal insulation of the crust. The larger the body, the higher the amount of energy stored as primordial heat, and the longer the time necessary to cool down the magma ocean. As previously explained, the flotation crust and the cumulate formed during the magma ocean episode are preserved during the solid-state regime that follows. Thus, following the magma ocean episode, planetesimals with radius $R>30$ km are likely to have formed a layered mantle. As illustrated in the example displayed in Figure \ref{fig:R270}, at the end of the magma ocean, a three-layer mantle exists and is composed of: (i) a flotation crust  composed of light crystals, (ii) a cumulate composed of heavy crystals, and (iii) a well-mixed interior that is a mixture of both types of crystals. The latter is the result of the incomplete segregation of crystals from the liquid suspension by the end of the magma ocean episode, i.e.: before the rheological transition.\\
\indent To complement this description, we show in Figure \ref{fig:solidification} the structure of the magma ocean at the end of its life, when the rheological transition occurs. We obtain that for small planetesimals, the mantle is  homogeneous. The larger the body, the thicker the crust and the cumulate and for planetesimals larger than $500\, \rm{km}$, crystal melt segregation is quasi-complete (over 80\%, Figure \ref{fig:solidification} (b)), and we get an``onion shell" structure. This two end-members scenario is consistent with $-$ and gives a physical explanation for $-$ classical models of magma ocean crystallization: (i) a homogeneous mantle, when crystals are assumed to stay in suspension during the evolution of the convective magma ocean \cite{Charlier18,Bryson19} and (ii) an "onion shell" mantle at the end of the magma ocean, when all crystals are assumed to settle as soon as they nucleate \cite{Maurice20}. Our model adds a trade-off between these two end-members, and allows for incomplete crystal segregation from the liquid suspension.

%
%

	\subsection{Influence of crystals properties}

	\begin{figure}
		\centering
		\includegraphics[width=0.45\textwidth]{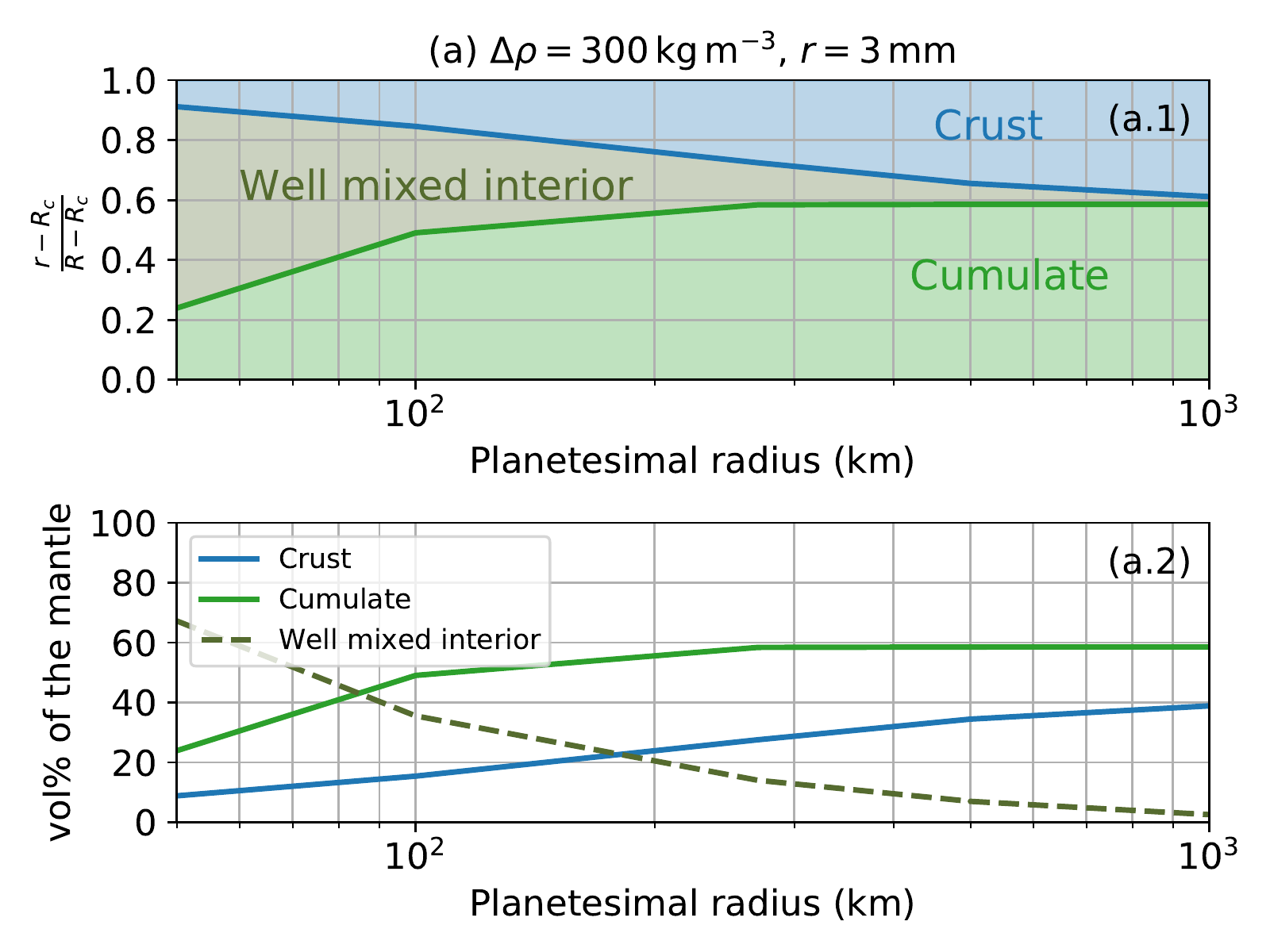}
		\includegraphics[width=0.45\textwidth]{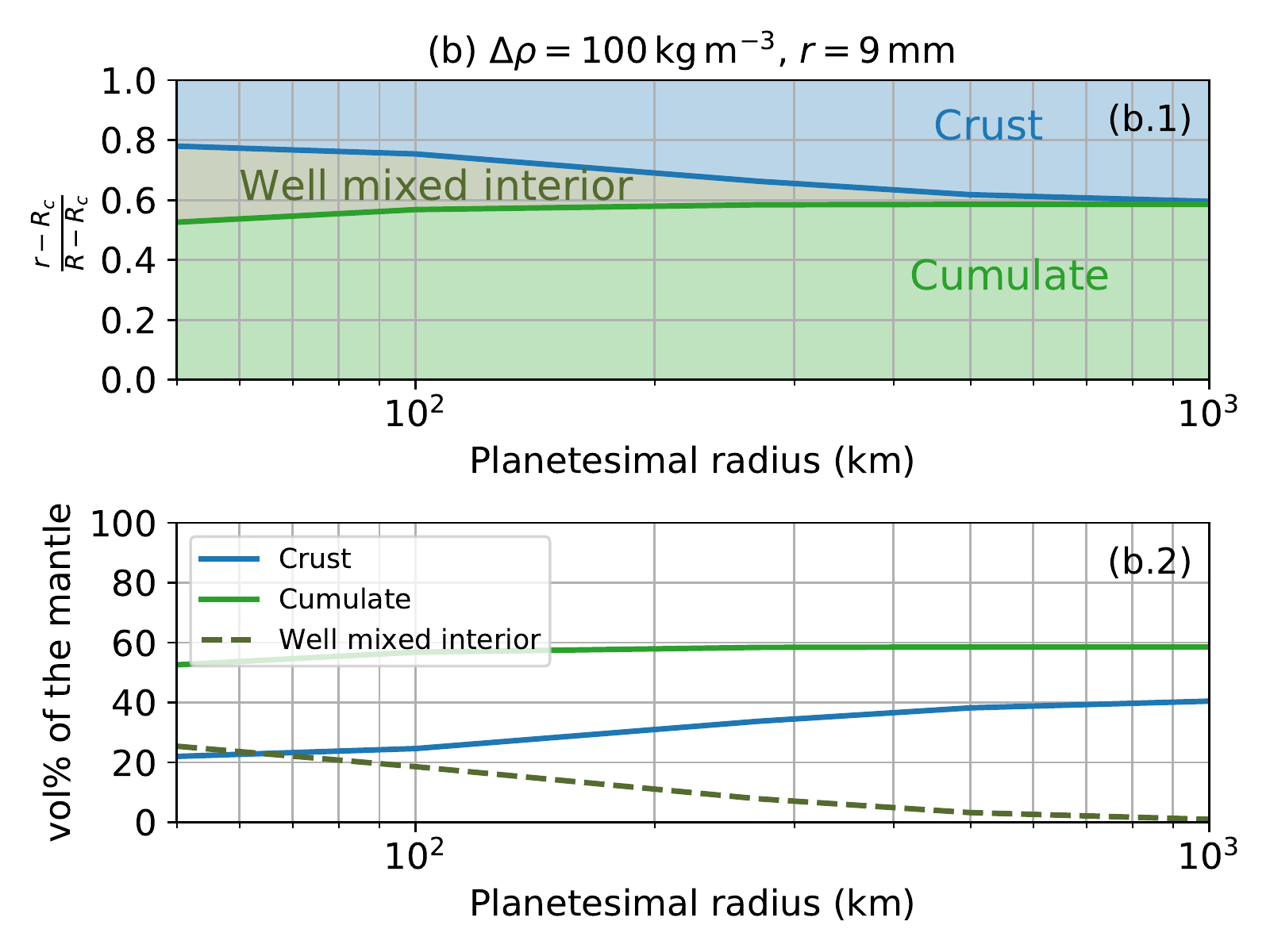}
		\caption{Influence of the crystal radius $r$ and the crystal/melt density difference $\Delta \rho$ on the internal structure of the planetesimal's mantle at the end of the magma ocean episode as a function of its radius. Two set of parameters are tested: (a) $\Delta \rho=300\ \rm{kg\, m^{-3}}$, $r=3\, \rm{mm}$, and (b) $\Delta \rho=100\ \rm{kg\, m^{-3}}$, $r=9\, \rm{mm}$. In both cases, we display the dimensionless thickness of each layer ((a.1) and (b.1)), and the corresponding fraction of the mantle occupied by each layer ((a.2) and (b.2)).}
		\label{fig:sensitivity_vesta}
	\end{figure}

	\begin{figure}
		\centering
		\includegraphics[width=\textwidth]{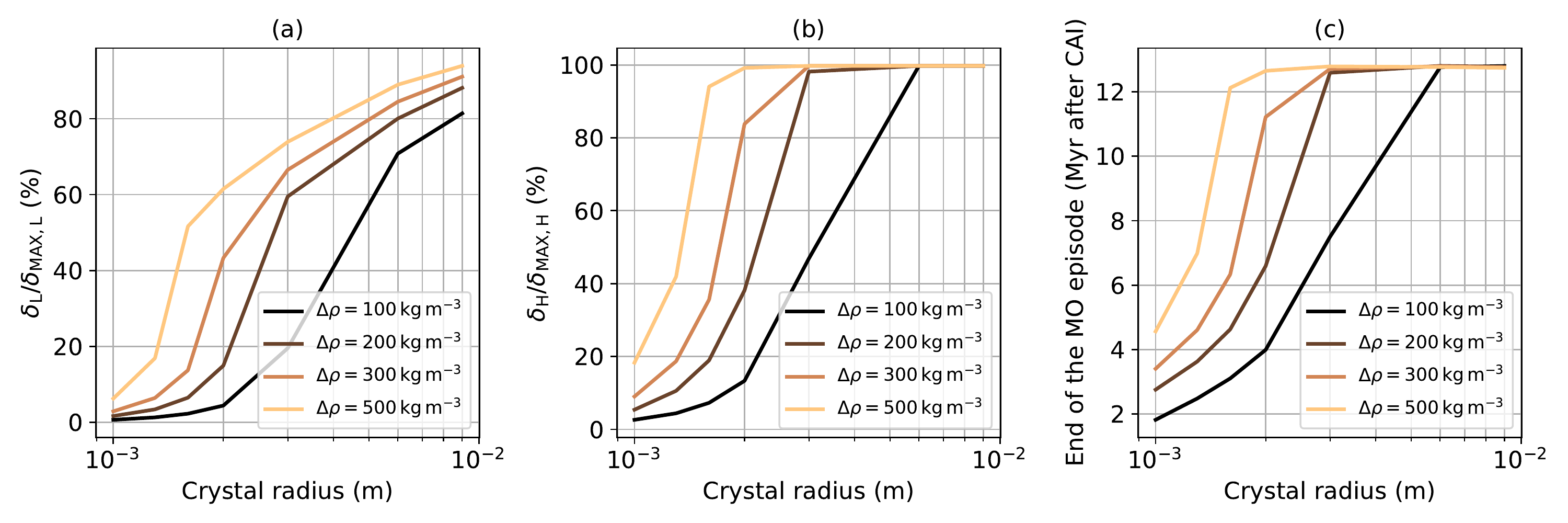}
		\caption{Influence of the crystal radius $r$ and the crystal/melt density difference on the formation of: the flotation crust (a), the cumulate (b) and the time at which the magma ocean ends (c) for a 270 km size planetesimal. $\delta_i$ is the thickness of the deposit $i$ at the end of the magma ocean, and $\delta_{MAX,i}$ is the maximum thickness that this deposit can reach once all crystals that compose $i$ have settled.}
		\label{fig:sensitivity_R}
	\end{figure}

\indent The stratification shown in Figure \ref{fig:solidification}  actually depends not only on the planetesimal radius, but also on the properties of the crystals that controls the segregation process. More precisely, and by definition of the settling velocity (\ref{eq:vs1})-(\ref{eq:vs2}), two parameters have a strong influence on the segregation processes: the crystal/melt density difference $\Delta \rho$ and the crystal radius $r$. For $\Delta \rho$, we choose in this study a reference value of $100\, \rm{kg\, m^{-3}}$. The exact value will depend on the crystal and the magma considered. It spans over $100-300\, \rm{kg\, m^{-3}}$ for plagioclase \cite{Namur11,ET12},  $150-400\, \rm{kg\, m^{-3}}$ for olivine and  $50-100\, \rm{kg\, m^{-3}}$ for pyroxene and orthopyroxene \cite{Suckale12b}. The radius of crystals that nucleate in the magma ocean is more critical although less constrained, but models for nucleation of crystals constrain $r$ between 1-10 mm \cite{Solomatov00}. \\
\indent To illustrate the effect of a change in these two properties on the layered structure of the mantle, we study the evolution of the internal structure when $\Delta \rho$ and $r$ are changed by a factor 3. Results are shown in Figure \ref{fig:sensitivity_vesta}. Small crystals induce a homogenized mantle, whereas large crystals are able to facilitate the formation of an onion-shell internal structure. Either for $\Delta \rho$ and for $r$, the increase induces a more efficient crystal/melt segregation and hence favors the onion-shell structure. Furthermore, we can note that the model is more sensitive to the crystal radius than to the relative density anomaly between the crystal and the liquid. This is consistent with the dependence of the deposit velocity: $v_s\sim \Delta \rho r^2$. \\
\indent In our toy-model, if all crystals settle, the planetesimal would be composed of a core representing 44\% of $R$, a cumulate at the CMB representing $33\%$ of $R$ and the last uppermost 23\% consisting in a flotation crust. The final thickness reached by the flotation crust $\delta_{\rm{L}}$ and the cumulate $\delta_{\rm{H}}$ at the end of the magma ocean episode can be smaller or equal to $\delta_{\rm{MAX,L}}$ and $\delta_{\rm{MAX,H}}$, the maximal thickness of the floating crust and of the cumulate, respectively. $\delta_{\rm{MAX,L}}$ and $\delta_{\rm{MAX,H}}$ are functions of the composition of the silicate phase and the radius of the planetesimal:
\begin{linenomath*}
	\begin{eqnarray}
		\frac{\delta_{\rm{MAX,L}}}{R}&=&1-\left(1-\phi_{\rm{L},0}\,\left[1-f_c^3\right]\right)^{1/3},\\
		\frac{\delta_{\rm{MAX,H}}}{R}&=&f_c\, \left(  \left[\phi_{\rm{H},0}\, \frac{1-f_c^3}{f_c^3}+1. \right]^{1/3}-1\right)
	\end{eqnarray}
\end{linenomath*}
\indent In Figure \ref{fig:sensitivity_R} (a) and (b), we study the thicknesses of each layers at the end of the magma ocean as a function of $\Delta \rho$ and $r$ for a 270 km radius planetesimal. The formation of the crust and the cumulate are more efficient if crystals are large and if the drop of density is important. Moreover, one can notice that the cumulate reaches its maximal thickness for crystal radii $r>3$ mm if $\Delta \rho \ge 200\, \rm{kg\, m^{-3}}$. Meanwhile, the flotation crust barely reaches 80\% of its maximal value for the largest crystals ($r=10$ mm). This asymmetry can be explained by the fact that the temperature of the magma ocean never becomes larger than the liquidus of the light crystals(1605$\rm{^oC}$), which is itself below the liquidus temperature of the heavy crystals (1800$\rm{^oC}$).  Therefore, heavy crystals can settle without being remelted. \\
\indent The efficiency of crystal segregation from the magma has also an influence on the duration of the magma ocean episode. The more efficiently crystals leave the magma ocean, the smaller the volume fraction of crystals remaining in suspension, and thus, the later the magma ocean reaches the rheological transition. This effect is illustrated in Figure \ref{fig:sensitivity_R} (c). In the limit of low $r$/low $\Delta \rho$, the segregation is inefficient and the rheological transition is reached early (down to 1 Myr). At large $r$/high $\Delta \rho$, the magma ocean lasts for more than 10 Myr due to a delayed rheological transition. This time is the one required for complete crystallization and differentiation of the magma ocean. In that case there is no well-mixed mantle encapsulated within the crust and the basal cumulate.

%
%

\section{Geophysical implications - the case of Vesta}

%
%

	\subsection{Layered structure of Vesta}
	
	\indent The asteroid 4 Vesta is a 267 km radius rocky differentiated body orbiting in the asteroids belt, and is considered as an almost intact pristine protoplanet \cite{Consolmagno15}. Vesta is thought to be the parent body of over 2000 achondrite meteorites called the HED series (howardite-eucrite-diogenite)  \cite{McSween13}. Eucrites are igneous rocks, mostly basaltic and gabbroic, whereas diogenites are cumulates rocks composed of pyroxenites and harzburgites and probably formed by fractional crystallization \cite{Zuber11}. Howardites are  brecciated basalts and pyroxenites, and correspond compositionally to a mixture of eucrites and diogenites. Meteoritic data indicate that the eucrite/diogenite mass ratio is 2:1 \cite{McSween13,McSween19}. \\
	\indent Crystallization models predict a layered structure of Vesta with an upper eucritic crust and a basal dunitic/harzburgitic cumulate. Depending on the model considered, diogenites could either result from the late crystallization of the remaining magma ocean between the eucritic crust and the cumulate \cite{Righter97,Neumann14}, or could corresponds to plutonic layers in the eucritic crust \cite{Mandler13}. These  predictions have been compared to the observation of the DAWN mission \cite{Russel13}. Two major craters (Rheasilvia and Veneneia) are located in the southern hemisphere \cite{Jaumann12,Marchi12}. Impact models predict that the impacts that form these craters induced an excavation depth up to 100 km \cite{Jutzi13}. The Vestean mantle should have been exposed, and according to previous crystallization models, diogenites and even olivine-rich cumulates. However, DAWN data only identified eucritic and howarditic material at the surface \cite{DeSanctis12}, and diogenites are not clearly detected \cite{McSween19}. Several hypotheses have been formulated to explain the inconsistency between petrological models and DAWN observations. First, it has been pointed out that the eucritic crust might be potentially very thick (up to 100 km), explaining why excavations did not unveil diogenites. But diogenite are however potentially present as plutonic patches and could correspond to density anomalies measured during the DAWN mission \cite{Mandler13,Clenet14,McSween19}. Another hypothesis is to rely on the mechanical mixing associated with impacts that enables the homogenization of the outer 100 km of Vesta  \cite{Jutzi13}. We can use our model to revisit this predicament.\\
	\indent In the previous section, we described in details the evolution of a planetesimal with a radius of 270 km which provies a scenario for the petrological evolution of Vesta in the hypothesis it accreted early (Figure \ref{fig:R270}). Magma ocean starts at 0.3 Myr, enabling the formation of an eucritic flotation crust and a olivine rich basal cumulate. The rheological transition is reached before the petrological evolution of the magma ocean has been completed. Thus, the transition to solid-state convection occurs in a 100 km thick well mixed mantle, relic of the magma ocean, encapsulated within a 10 km thick eucritic upper crust and a 30 km thick olivine rich basal cumulate. As illustrated in Figure \ref{fig:R270} (d), the well-mixed mantle is composed of light and heavy crystals in a ratio close to 1:1, corresponding to an howarditic composition. Hence, if an impact excavates material 100 km depth, not only would the 10 km thick eucritic crust be removed \cite{Jutzi13} , but the howarditic mantle would also be exposed. Hence the incomplet petrological evolution of the Vestean magma ocean provides a consistent scenario to interpret DAWN observations.

%
%

\subsection{Vestean thermal evolution}

	\begin{figure}
		\centering
		\includegraphics[width=\textwidth]{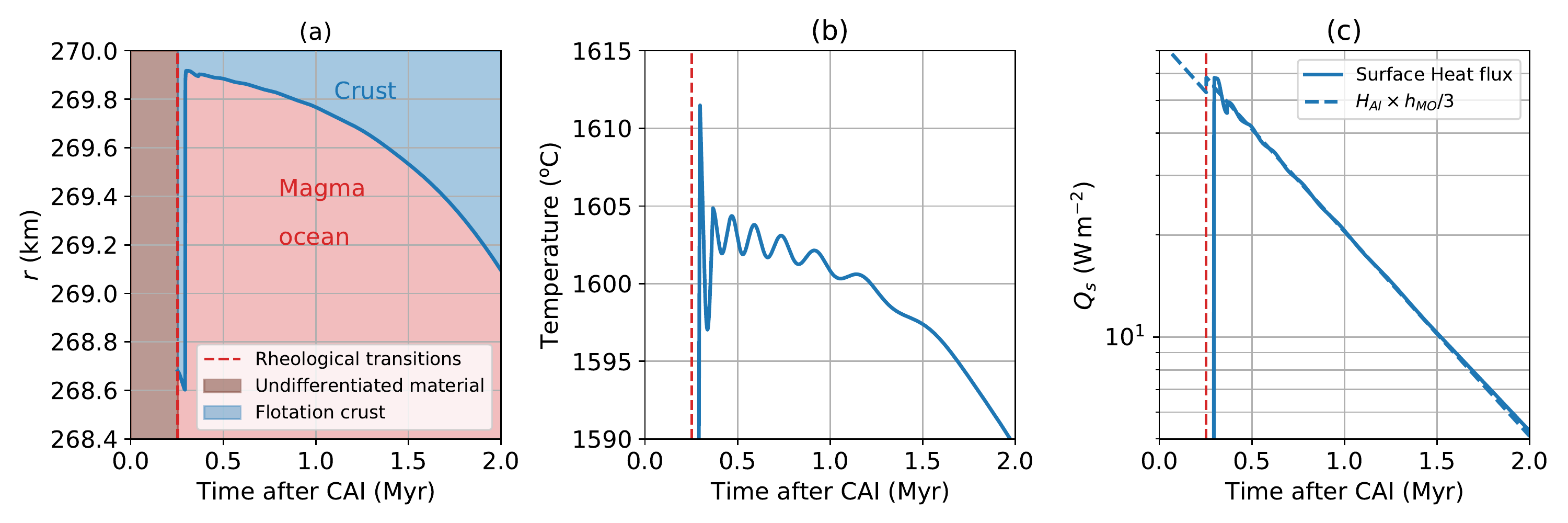}
		\caption{Thermal scenario for Vesta focusing on the remelting episodes that occur soon after the onset of the magma ocean. (a) Evolution of the flotation crust thickness. The surface corresponds to $r=270\, \rm{km}$. (b) Evolution of the magma ocean temperature. (c) Evolution of the surface heat flux and the heat generated by $^{26}Al$ in the magma ocean.}
		\label{fig:remelting}
	\end{figure}
	
	\indent Additional insights on the thermal evolution of Vesta can be grained from the HED meteorites analyses. In particular, rare-Earth elements (REE). REE abundances in diogenites \cite{Barrat08} and eucrites \cite{Yamaguchi09} in Vestean meteorites show that they are not co-genetic and suggest that the formation of diogenites and eucrites on Vesta results from a non-monotonic thermal history involving remelting episodes of previously crystallized cumulates. Usually, external energy sources are called for to produce the remelting episodes, such as large impactors. However, our toy-model proposes an intrinsic mechanism that provides new insights in these processes.\\
	\indent In an early accreted Vesta, the heat generated by the radioactive decay of $\rm{^{26}Al}$ is strong enough to enable high degree of melting. However, as shown in Figure \ref{fig:CondConv}, the temperature does not go above $\approx$1600-1610 $\rm{^oC}$. To understand this effect, we focus on the first 3 Myr, and we display the evolution of the magma ocean temperature and the surface heat flux in Figure \ref{fig:remelting}. After the rheological transition has been reached, the crust is initially 1.25 km thick, the thickness of the stagnant lid at the transition. At this time, as discussed previously, flotation of light crystals occurs and the crust thickens. Because of the insulating effect of the crust, the temperature of the magma ocean increases until it reaches the liquidus temperature of the crust ($T_l^{L}=1605\rm{^oC}$). This induces a melting of the base of the crust and a crustal thinning until the temperature at its base decreases to $T_l^{L}$. As the crust thins, the surface heat flux increases, and becomes larger than the rate of internal heat generation (Figure \ref{fig:remelting} (c)). The temperature of the magma ocean thus decreases and goes beyond the light crystals liquidus. This triggers light crystal nucleation and flotation, hence a thickening of the crust.  Because of the insulation effect of the crust, the heat flux decreases, and a new unbalance between heat losses and heat generation induces a temperature increase in the magma ocean. As a consequence, the temperature rises again, and a new remelting occurs. Hence, our model provides an intrinsic mechanism that enables cycles of remelting-crystallization episodes, characterized by a wobbling temperature with an amplitude of 15-20$\rm{^oC}$. The crust formed in these conditions will display the petrological record of this non-monotonic thermal history. Although discussed here in the case of Vesta, the wobbling thermal history is an intrinsic feature of our model and could be used to interpret other families of meteorites, such as pallasites that also recorded remelting episodes \cite{Barrat21}.

%
%

\begin{figure}
	\centering
	\includegraphics[width=0.8\textwidth]{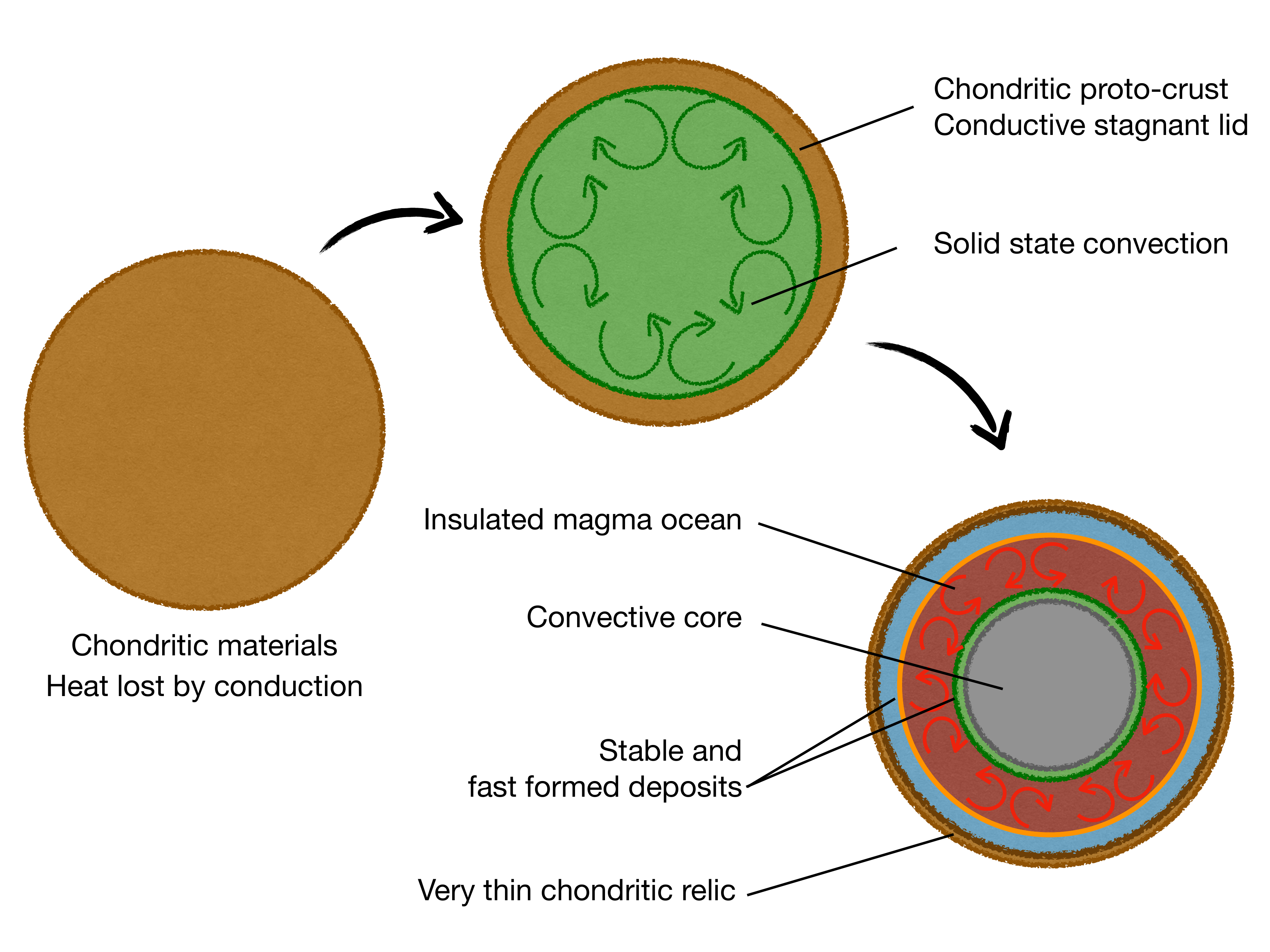}
	\caption{General evolution of a planetesimal with a radius larger than 30 km, instantaneously accreted during the first few Myr of the solar system. The undifferentiated planetesimal is heated by the radioactive elements, which triggers solid-state convection. Convection occurs under a stagnant lid of undisrupted material at the surface. Core differentiation follows partial melting that further induces a rheological transition to a magma ocean episode in the proto-mantle when it reaches $40\rm{vol}\%$ of melt. During the magma ocean episode, a basal cumulate and a flotation crust are produced. The insulation effect of the crust on the magma ocean increases the duration of the episode from 1 to 100 Myr $-$ depending on the planetesimal radius. Intrinsic remelting cycles occur at the base of the crust during meanwhile. The basal cumulate and the flotation crust generally encapsulate a well-mixed mantle.}
	\label{fig:FigG}
\end{figure}

\section{Conclusion}

\indent We develop a model for erosion/deposition of an erodible bed, and we adapt the Shields' formalism to study the evolution of magma oceans in planetary systems. The model is used to establish a generic scenario for early accreted planetesimals, as illustrated in Figure \ref{fig:FigG}. To form basal cumulates and/or flotation crust, planetesimals must have undergone an episode of magma ocean, which happens for planetesimal radius larger than $R>30$ km provided they accreted early. Beyond the rheological transition that marks the transition to the magma ocean episode, the mantle is a liquid suspension that convects under the relics of the stagnant lid which has become a km-size proto-crust. Crustal remelting/flotation cycles happen at this time, modifying the crust thickness and maintaining the magma ocean temperature at moderate (sub-liquidus) values.  After a 1 to 100 Myr magma ocean episode, depending on the planetesimal size, the mantle reaches the rheological transition again and transitions back to a solid-state convection regime. Consequently, deposits that have formed by crystal segregation during the magma ocean episode become unstable. However, both the basal cumulate and the flotation crust are kinetically preserved as the erosion time scale is very large. Thus the deposits preserves the specific informations (e.g.: remelting episodes, magnetization,...) they have stored.\\
\indent The model could be adapted to study magma ocean formed in larger bodies and produced by giant impacts. The layered structure produced during the magma ocean episode sets the initial conditions for the solid state convection in planetary mantle. The relics of this structure could be related to preserved heterogeneities e.g.: in the Earth (Large low-shear-velocity provinces or Ultra low velocity zones) or in the Moon (KREEP crust).

%

\section*{Declaration of interest}
The authors report no conflicts of interest.
		
%
%

\acknowledgments
\indent This paper is part of Cyril Sturtz's PhD thesis (Universit\'e de Paris, Institut de Physique du Globe de Paris). The authors wish to thank Marc Chaussidon for fruitful discussions and comments. This study contributes to the IdEx Universit\'e de Paris ANR-18-IDEX-0001. This work was supported by the Programme National de Plan\'etologie (PNP) of CNRS/INSU, co-funded by CNES.

%
%

\appendix

%
%

\section{Convection under a stagnant lid}
\label{App:SLC}

\begin{figure}
	\centering
	\includegraphics[width=0.45\textwidth]{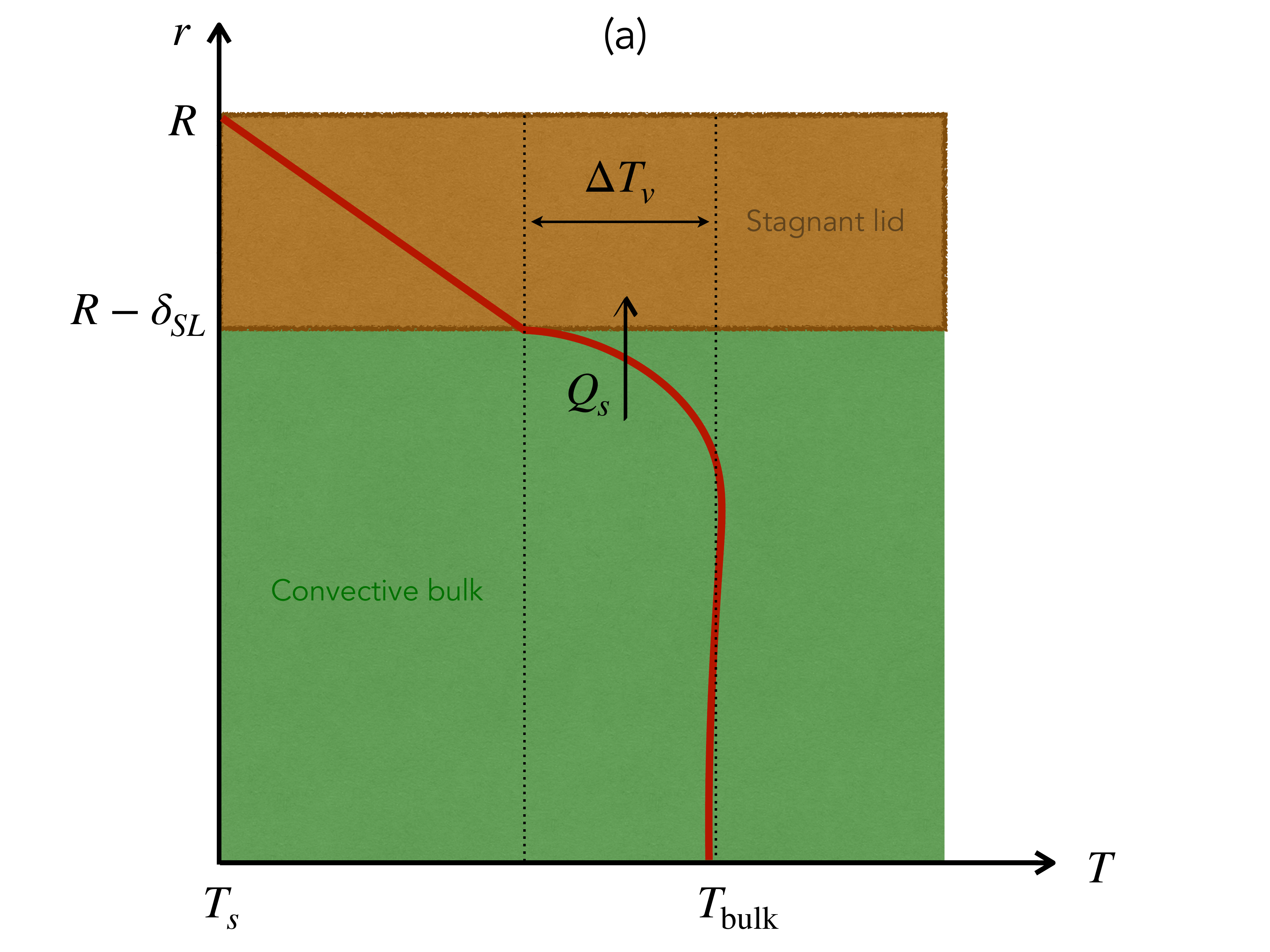}
	\includegraphics[width=0.7\textwidth]{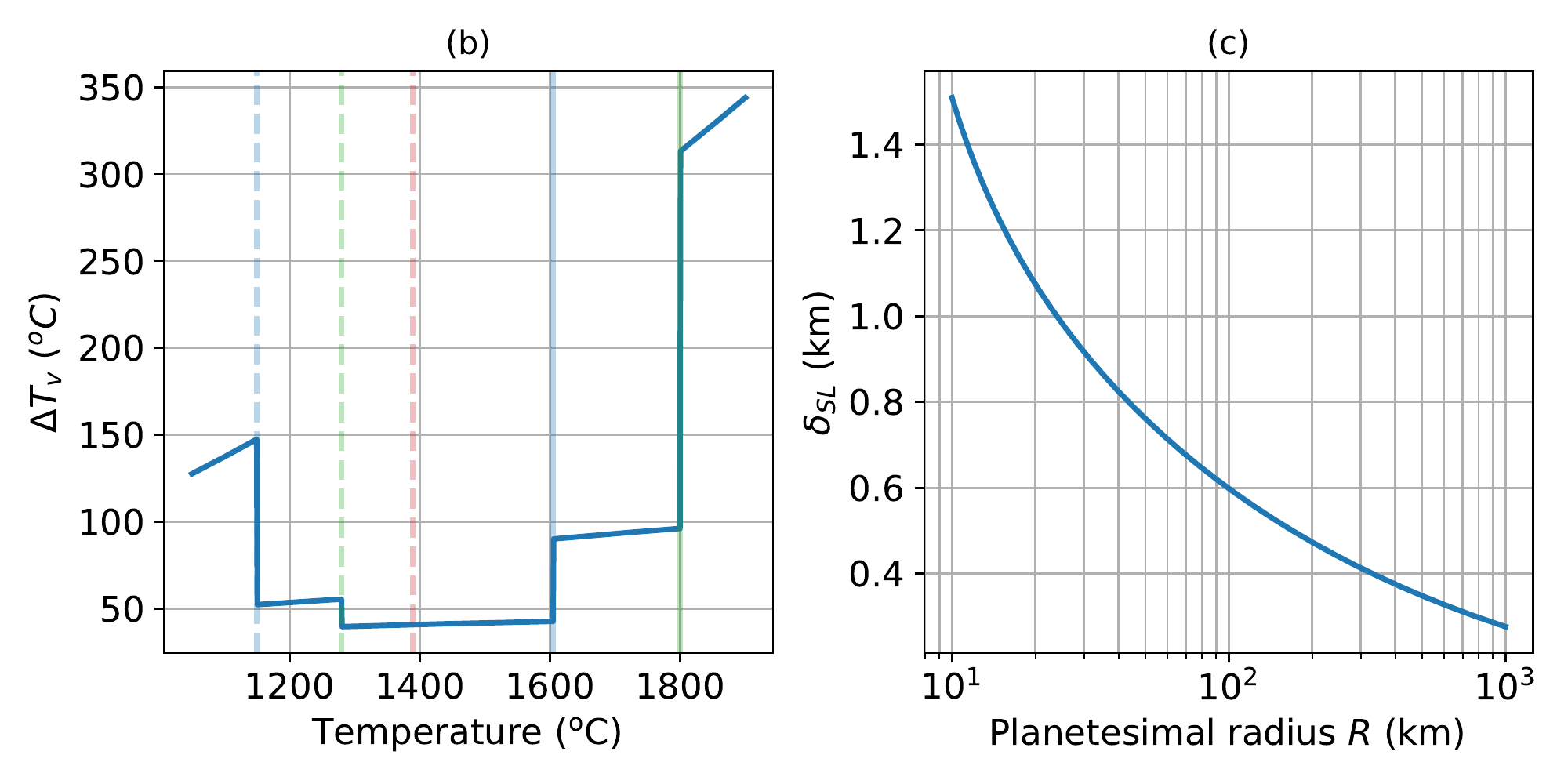}
	\caption{(a) Schematic view of the convective system under a stagnant lid. (b) Calculation of the temperature drop due to viscous effects as a function of temperature. (c) Stagnant lid thickness at the rheological transition as a function of the planetesimals' radius.}
	\label{fig:SL}
\end{figure}
\begin{figure}
	\centering
	\includegraphics[width=\textwidth]{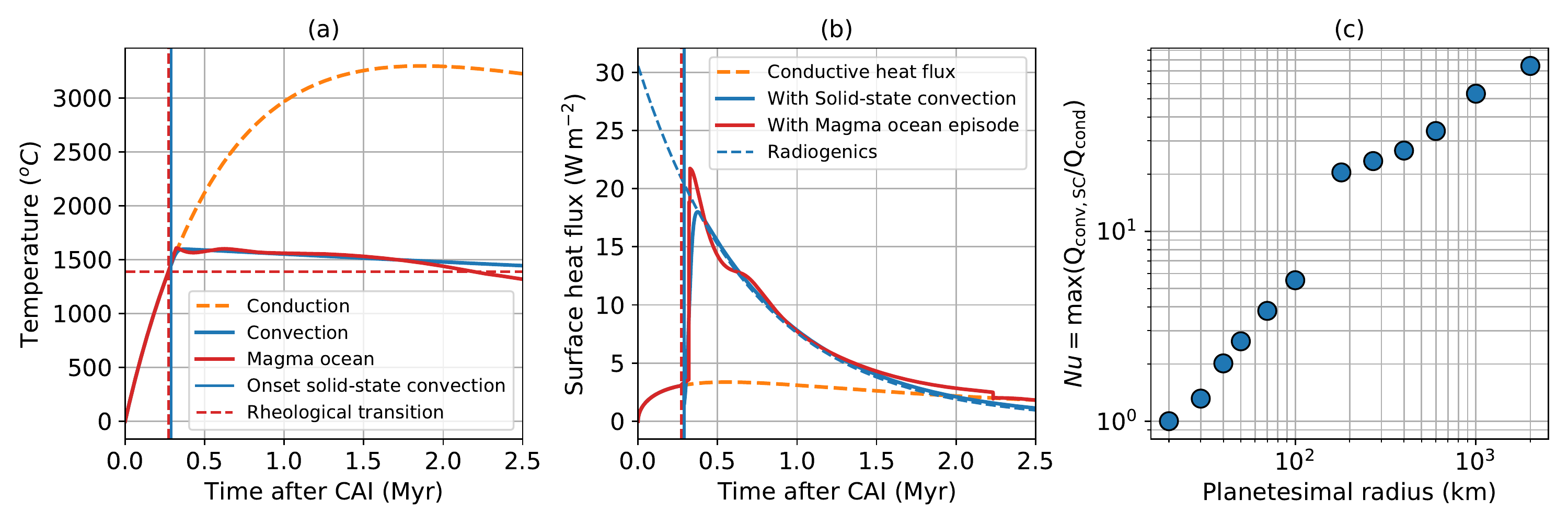}
	\caption{Evolution of the temperature (a) and of the heat flux (b) in the case of a 180 km radius planetesimal that accreted at $t=0$ Myr in the case of purely conductive regime (orange), with the solid-state convection episode (blue) and including the magma ocean episode (red). (c) Comparison between the conductive and the convective (in solid state convection) heat flux, thanks to the Nusselt number defined as the maximal value of the ratio between both fluxes in terms of Nusselt number.}
	\label{fig:AppCondConv}
\end{figure}
\indent As the surface temperature and the bulk temperature are far one from the other, the viscosity varies on several orders of magnitude. \citeA{Davaille93} proved that in this case the convection occurs under a stagnant lid where heat is transported by conduction (see Figure \ref{fig:SL} (a)). In this way, the characteristic temperature scale is given by:
\begin{linenomath*}
	\begin{equation}
		\Delta T_v=-C_v\, \frac{\eta}{\frac{\rm{d}\eta}{\rm{d}T}},
	\end{equation}
	\end{linenomath*}
with $C_v=2.24$. Assuming the rheology (\ref{eq:visco_SC1}-\ref{eq:visco_SC2}), using also (\ref{eq:phi}), we get :
\begin{linenomath*}
	\begin{eqnarray}
		\Delta T_v(T)&=&C_v\, \left[\sigma_{Sil}\frac{\partial \phi}{\partial T}+ \frac{E_{a,Sil}}{R_gT^2}\right]^{-1},\\
		\frac{\partial \phi}{\partial T}&=&\epsilon_{L}\frac{\phi_{Euc,0}}{T_{l}^{Euc}-T_s^{Euc}}+\epsilon_{H}\frac{\phi_{Oli,0}}{T_{l}^{Oli}-T_s^{Oli}},
	\end{eqnarray}
\end{linenomath*}
with $\epsilon_i=1$ if $T_l^i>T>T_s^i$, and 0 otherwise. $\Delta T_v$ stands for the drop of temperature in the convective bulk, and is plotted in Figure \ref{fig:SL} (b). The basal flux entering the stagnant lid is imposed by convection, and the link between the stagnant lid thickness and the basal heat flux is given by Fourier's law in spherical coordinates:
\begin{linenomath*}
	\begin{equation}
		Q_s=\lambda_{Sil}\, \frac{T_{\rm{bulk}}-\Delta T_v -T_s}{\delta_{SL}}\frac{R}{R-\delta_{SL}},
	\end{equation}
\end{linenomath*}
where $Q_s$ is the heat flux given by the scaling law for convection:
\begin{linenomath*}
	\begin{equation}
		Q_s=\lambda_{Sil}\, \left(\frac{\alpha_{Sil}\rho_{Sil}g}{\kappa_{Sil}\eta(T)}\right)^{1/3}\, \left(\frac{\Delta T_v}{C_T}\right)^{4/3}.
	\end{equation}
\end{linenomath*}
It leads to the expression of the stagnant lid thickness:
\begin{linenomath*}
	\begin{equation}
		\delta_{SL}(T)=\frac{R}{2}\, \left(1- \sqrt{1-\frac{4 \lambda (T-\Delta T_v-T_s)}{Q_s R}}\right),
	\end{equation}
\end{linenomath*}
Thus, the stagnant lid thickness is a function of the temperature and the planetesimal's radius. We can estimate the thickness of the proto-crust at the onset of magma ocean by calculating $\delta_{SL}(T=T_{RT},R)$. Results are plotted in Figure \ref{fig:SL} (c), and we can see that the stagnant lid is km-thick when magma ocean triggers. \\
\indent For the onset of convection, the Rayleigh-Roberts number has to be grater than the critical value of $Ra_{H,c}=5758$. It has also been pointed out that this onset is delayed due to the strong temperature dependent viscosity \cite{Choblet00}. The delay time $\tau_{visc}$ is given by the relationship:
\begin{linenomath*}
	\begin{equation}
		\tau_{visc}=\frac{1}{\pi}\, \frac{R^2}{\kappa}\, \left(\frac{Ra_{H,c}}{Ra_H}\right)^{2/3}\, \left(\frac{T-T_s}{\Delta T_v}\right)^{8/3}
	\end{equation}
\end{linenomath*}
In this framework, we compare the thermal evolution of a 180 km size planetesimal in both the case of fully conductive life and by taking into account solid-state convection. The evolutions of the bulk temperature and the surface heat flux in both cases are illustrated in Figure \ref{fig:AppCondConv} (a) and (b). The thermal evolution shows that the thermal history experiences a maximum in both case. But the convection attenuates this maximum, as discussed in the paper and by \citeA{Kaminski20}. This attenuation is explained by the rise of the heat flux in case of convection, compared to the surface heat flux in case of conduction. To quantify this effect, we can define a modified Nusselt number, as the maximal value of the ratio between the convective heat flux and the conductive one. Results are displayed in Figure \ref{fig:AppCondConv} (c), and highlights a rise of heat losses that increases with the planetesimal radius.

\section{Crystals that remains in suspension in the magma ocean}
\label{App:MassCons}
We consider a planetesimal of radius $R$ with an iron core of radius $R_c$. The magma ocean is at temperature $T$ and the volume of light and heavy crystals that crystallise are given respectively by:
\begin{linenomath*}
	\begin{eqnarray}
		V^L_{\rm{cr}}&=&\frac{4}{3}\pi(R^3-R_c^3)\, \phi_{L,\rm{cr}},\\
		V^H_{\rm{cr}}&=&\frac{4}{3}\pi(R^3-R_c^3)\, \phi_{H,\rm{cr}},
	\end{eqnarray}
\end{linenomath*}
with $\phi_{i,\rm{cr}}$ given by (\ref{eq:phiLcr}) and (\ref{eq:phiHcr}). The magma ocean is encapsulated between a crust of thickness $\delta_L$ composed of light crystals, and a cumulate of thickness $\delta_H$ composed of heavy one. The volume occupied by the crust and the cumualte are respectively given by:
\begin{linenomath*}
	\begin{eqnarray}
		V^L_{\rm{dep}}&=&\frac{4}{3}\pi [R^3-(R-\delta_L)^3],\\
		V^H_{\rm{dep}}&=&\frac{4}{3}\pi [(R_c+\delta_H)^3-R_c^3].
	\end{eqnarray}
\end{linenomath*}
As the volume of the magma ocean is:
\begin{linenomath*}
	\begin{eqnarray}
		V_{\rm{MO}}&=&\frac{4}{3}\pi [(R-\delta_L)^3-(R_c+\delta_H)^3],
	\end{eqnarray}
\end{linenomath*}
the fraction of light and heavy crystal in suspension are given by $\phi_{i,\rm{sus}}=(V^i_{\rm{cr}}-V^i_{\rm{dep}})/V_{\rm{MO}}$, so that:
\begin{linenomath*}
	\begin{eqnarray}
		\phi_{L,\rm{sus}}&=&\frac{  (1-f_c^3)\phi_{L,\rm{cr}}  - 1 + (1-\delta_L/R)^3   }{(1-\delta_L/R)^3-(f_c+\delta_H/R)},\\
		\phi_{H,\rm{sus}}&=&\frac{  (1-f_c^3)\phi_{H,\rm{cr}}  - (f_c+\delta_H/R)^3 + f_c^3   }{(1-\delta_L/R)^3-(f_c+\delta_H/R)}.
	\end{eqnarray}
\end{linenomath*}

\bibliography{biblio}

\end{document}